\documentclass[%
 reprint,
 showkeys,
 floatfix,
 amsmath,amssymb,
 aps,
 prd,
]{revtex4-2}

\usepackage{graphicx}% Include figure files
\usepackage{dcolumn}% Align table columns on decimal point
\usepackage{bm}% bold math
\usepackage{aas_macros}
\usepackage{xcolor}
\definecolor{xlinkcolor}{cmyk}{1,1,0,0}

\usepackage{siunitx}
\PassOptionsToPackage{hyphens}{url}
\usepackage[bookmarks=true,colorlinks=true,]{hyperref}
\begin{document}

\preprint{APS/RotCCSNwPT}

\title{Impact of Rotation on the Multimessenger Signatures of a Hadron-quark Phase Transition in Core-collapse Supernovae}% Force line breaks with \\

\author{Shuai Zha}
\email{szha.astrop@gmail.com}
\affiliation{Tsung-Dao Lee Institute, Shanghai Jiao Tong University, Shanghai 200240, China}

\author{Evan O'Connor}
\affiliation{The Oskar Klein Centre, Department of Astronomy, Stockholm University, AlbaNova, SE-106 91 Stockholm, Sweden}

\date{\today}% It is always \today, today,

\begin{abstract}
We study the impact of rotation on the multimessenger signals of core-collapse supernovae (CCSNe) with the occurrence of a first-order hadron-quark phase transition (HQPT). We simulate CCSNe with the \texttt{FLASH} code starting from a 20~$M_\odot$ progenitor with different rotation rates, and using the RDF equation of state from \textit{Bastian} 2021 that prescribes the HQPT. Rotation is found to delay the onset of the HQPT and the resulting dynamical collapse of the protocompact star (PCS) due to the centrifugal support. All models with the HQPT experience a second bounce shock which leads to a successful explosion. The oblate PCS as deformed by rotation gives rise to strong gravitational-wave (GW) emission around the second bounce with a peak amplitude larger by a factor of $\sim10$ than that around the first bounce. The breakout of the second bounce shock at the neutrinosphere produces a $\bar{\nu}_e$-rich neutrino burst with a luminosity of serveral 10$^{53}$~erg~s$^{-1}$. In rapidly rotating models the PCS pulsation following the second bounce generates oscillations in the neutrino signal after the burst. In the fastest rotating model with the HQPT, a clear correlation is found between the oscillations in the GW and neutrino signals immediately after the second bounce. In addition, the HQPT-induced collapse leads to a jump in the ratio of rotational kinetic energy to gravitational energy ($\beta$) of the PCS, for which persistent GW emission may arise due to secular nonaxisymmetric instabilities. 
\end{abstract}

%\keywords{Core-collapse supernova}%Use showkeys class option if keyword

\maketitle

%\tableofcontents

\section{Introduction} \label{sec:intro}
The protocompact star (PCS) born in a core-collapse supernova (CCSN) is an exceptional laboratory for high-density nuclear physics. Inside a PCS, matter can reach densities higher than a few times the nuclear saturation density ($\rho_{\rm sat}\simeq2.7\times10^{14}$\,g\,cm$^{-3}$),  temperatures of $k_{\rm B}T\sim$10-100\,MeV, and the proton to baryon ratio $Y_p$ of $\sim 0.2$ or lower. These extreme conditions can hardly be accessible simultaneously in terrestrial experiments. Such a site may allow the dissociation of hadrons such as protons and neutrons into quarks, namely, a hadron-quark phase transition (HQPT) \cite{1984PhRvD..30..272W,1991Sci...251.1194O,2022arXiv220317188B}. The occurrence of a HQPT can lead to the collapse of a PCS and induce a subsequent bounce shock due to the stiffening of the quark matter equation of sate (EOS). This provides an alternative plausible mechanism for CCSN explosions \cite{1988ApJ...335..301T,1993ApJ...414..701G,2009PhRvL.102h1101S,2018NatAs...2..980F,2022arXiv220410397J}, other than the canonical neutrino-driven and magneto-driven mechanisms \cite{1990RvMP...62..801B,2002RvMP...74.1015W,2012ARNPS..62..407J,2021Natur.589...29B}. 

However, it is a challenging task to probe the properties and dynamics of a PCS, being buried in the stellar envelope that is opaque to electromagnetic waves. Modern messengers, such as gravitational waves (GWs) and neutrinos, can provide an unobstructed view of a PCS thanks to their feeble interactions with ordinary matter. For example, the detection of neutrinos from SN1987A \cite{1987PhRvL..58.1490H,1987PhRvL..58.1494B} has confirmed the basic scenario of stellar core collapse in Type II supernovae \cite{1987PhRvL..59..938B,1987Natur.327..597H,1990RvMP...62..801B}. It is proposed that both neutrino and GW signals can be used to trace the postbounce PCS contraction \cite[see eg.][]{2007PhR...442...38J,2013ApJ...762..126O,2019ARNPS..69..253M,2019PhRvL.123e1102T,2021ApJ...923..201E}. 

Recent studies with advanced supernova simulations have predicted unique neutrino \cite{2009PhRvL.102h1101S,2010ApJ...721.1284N,2000ApJS..131..273F,2018NatAs...2..980F,2021ApJ...911...74Z} and GW \cite{2020PhRvL.125e1102Z,2022ApJ...924...38K} signatures associated with the HQPT occurring in a CCSN. The shock breakout at the neutrinosphere following the bounce of the quark-matter core sends out a neutrino burst with a luminosity of several $10^{53}$\,erg\,s$^{-1}$ in a few milliseconds \cite{2009PhRvL.102h1101S,2018NatAs...2..980F}. The burst is rich in electron antineutrinos that are readily captured by large, water-based or liquid scintillator-based terrestrial detectors \cite{2003NIMPA.501..418F,2011A&A...535A.109A,2016JPhG...43c0401A} through the inverse $\beta$-decay process \cite{2012ARNPS..62...81S}. Meanwhile, 2D axisymmetric simulations of nonrotating CCSNe suggest that the dynamical collapse of a hadronic-matter core to a quark-matter core results in a loud kHz GW burst within a few milliseconds \cite{2020PhRvL.125e1102Z,2022ApJ...924...38K}. The amplitude of this burst is a few ten times that of the GW signals in normal CCSNe with nonrotating progenitors.

It is known that stars generally rotate to some extent. Massive presupernova stars with fast rotating iron cores are thought to be progenitors of hypernovae and long soft gamma-ray bursts \cite{1970AZh....47..813B}. Rotation can naturally deform stellar cores during their collapse and bounce, and thus enhance the GW emission of CCSNe in the corresponding episodes, as shown by perturbation analysis \cite{1977ApJ...214..566S,1979sgrr.work..383T,1974ApJ...191L.105T} and hydrodynamic simulations \cite{1982A&A...114...53M,1991A&A...246..417M,1997A&A...320..209Z,2002A&A...393..523D,2002ApJ...565..430F,2004ApJ...600..834O,2004PhRvD..69h4024S,2014PhRvD..90d4001A,2021ApJ...914...80P}. Moreover, it has been found that measuring the GW frequency and amplitude can provide the otherwise inaccessible information about the stellar core, such as its angular momentum and compactness \cite{2014PhRvD..90d4001A,2021ApJ...914...80P}, as well as constrain the nuclear matter EOS \cite{2017PhRvD..95f3019R,2021ApJ...923..201E}. 

Following the same principle, the dynamical collapse of a rotating PCS due to the HQPT in CCSNe can also emit characteristic GW signals, which is the major subject of this work. Previous studies have simulated the collapse of rotating neutron stars into quark stars and predicted the associated GW signals \cite{2006ApJ...639..382L,2009MNRAS.392...52A}. However, in \cite{2006ApJ...639..382L,2009MNRAS.392...52A} the initial rotating neutron stars are constructed in hydrostatic equilibrium with a polytropic EOS and the collapse is triggered by switching to a hybrid EOS that includes quarks. Strange quark matter with equal amounts of $u$, $d$ and $s$ quarks is assumed after the HQPT and it is described by the MIT bag model EOS \cite{1986A&A...160..121H}. The onset of the HQPT is parametrized with a transition density of $2.6\rho_{\rm sat}$ and pure quark matter appears above $9\rho_{\rm sat}$. We note that these transition densities are highly uncertain and the values used in \cite{2006ApJ...639..382L,2009MNRAS.392...52A} are one possible parametrization. These exploratory simulations predicted the emitted GW signals with frequencies of a few kHz and dimensionless strain amplitudes of $\mathcal{O}(10^{-22})$ ($\mathcal{O}(10^{-19})$) for a 10-Mpc (10-kpc) event. 

In this paper, we perform more realistic simulations of CCSNe with a HQPT starting from rotating massive stars. The HQPT is prescribed with a finite-temperature EOS from the set of \textit{Bastian} 2021 \cite{2021PhRvD.103b3001B} whose variants have been frequently used in simulations of CCSNe \cite{2018NatAs...2..980F,2020ApJ...894....9F,2022ApJ...924...38K,2022arXiv220410397J} and binary compact mergers \cite{2019PhRvL.122f1102B}. We study the impact of rotation on the postbounce dynamics and the characteristic GW and neutrino signals associated with the HQPT. With several different initial rotation rates, we roughly outline the quantitative relation between the signal strength and the progenitor rotation.

Our paper is organized as follows. In \S\ref{sec:method} we describe the methodology used in our simulations. Then we present our results on the CCSN dynamics (\S\ref{ssec:dyn}), GW signals (\S\ref{ssec:gw}) and neutrino signals (\S\ref{ssec:neu}). We give our conclusions in \S\ref{sec:conclu}.

\section{Methodology} \label{sec:method}
\subsection{Equation of state}
We use the RDF 1.9 EOS from \cite{2021PhRvD.103b3001B} that prescribes a first-order HQPT at supranuclear densities for the description of the stellar matter in CCSNe. The underlying model and parameters of this EOS are presented in \cite{2021PhRvD.103b3001B} and here only relevant information is given for the completeness. Homogeneous hadron and quark matter are described using the relativistic density functional formalism \cite{2017PhRvD..96e6024K}. The phase of hadronic matter is described with the DD2 EOS of \cite{2010NuPhA.837..210H} which employs the relativistic mean-filed approach with density-dependent meson-nucleon coupling \cite{1999NuPhA.656..331T} and DD2 parameterization \cite{2010PhRvC..81a5803T}. The phase of quark matter is described by the string-flip model for two-flavor ($ud$) quark matter \cite{2017PhRvD..96e6024K}. Vector repulsion is introduced at high densities to make the quark matter EOS stiff enough to allow for a 2-$M_\odot$ compact star \cite{2010Natur.467.1081D,2013Sci...340..448A}. The mixed hadron-quark phase during the HQPT is constructed with the Gibbs conditions for phase equilibrium (see, e.g. Chapter 3 of \cite{1969stph.book.....L}) and considering the chemical equilibrium for baryon and charge simultaneously \cite{1992PhRvD..46.1274G}. 

In this work, we choose the RDF EOS with the parameter set RDF 1.9 (cf. Table I of \cite{2021PhRvD.103b3001B}) which leads to a maximum mass of 2.17\,$M_\odot$ for cold compact stars. We note that the RDF EOSs with other parameter sets (RDF 1.1 $\ldots$ RDF 1.7) were previously used in simulations of CCSNe \cite{2018NatAs...2..980F,2022ApJ...924...38K,2022arXiv220410397J} and binary neutron-star mergers \cite{2019PhRvL.122f1102B}. Here our particular choice of RDF 1.9 is for the early onset of the HQPT at $\sim1.8\times\rho_{\rm sat}$ for zero temperature which corresponds to a cold compact-star mass of $\sim$0.81~$M_\odot$. This is pragmatic for our exploration to save computational time, particularly for rotating CCSNe. We do not argue the correctness of this choice but expect qualitatively similar results for other RDF EOSs in \cite{2021PhRvD.103b3001B}. Of course, the quantitative dependence on the EOS should be investigated in the future as has been done for compact object mergers \cite{2019PhRvL.122f1102B}.

\begin{figure}[t!]
    \centering
    \includegraphics[width=0.48\textwidth]{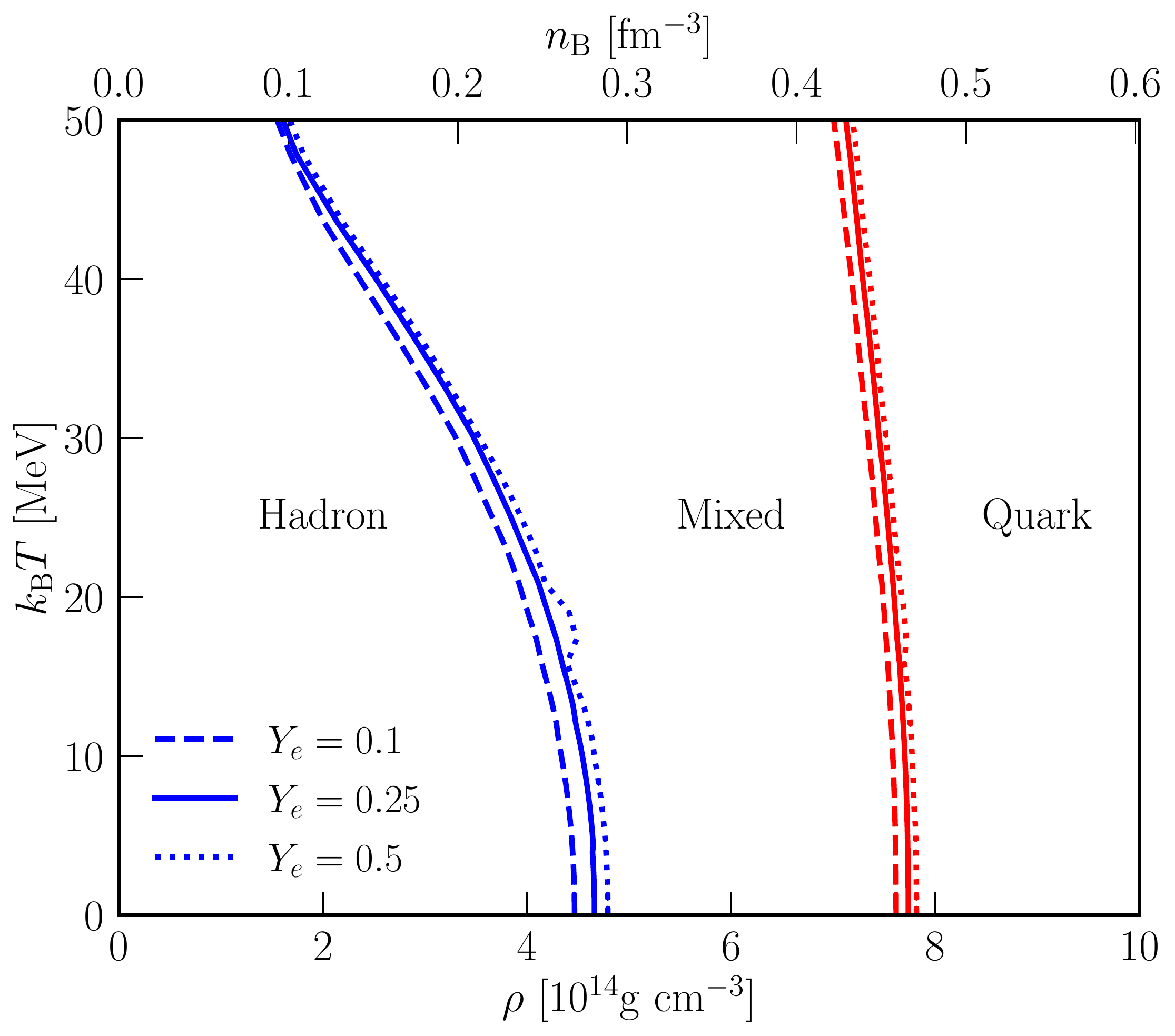}
    \caption{Hadron-quark phase diagram in the RDF 1.9 EOS. Blue lines are the boundaries of the hadronic and mixed phases and red lines are the boundaries of the mixed and quark phases. Dashed, solid and dotted lines correspond to the electron fraction $Y_e$ of 0.1, 0.25 and 0.5. The turning behavior for $Y_e=0.5$ and $k_{\rm B}T=15$-$25$~MeV is likely an artifact due to the tabulated EOS.}
    \label{fig:phase_eos}
\end{figure}

Figure~\ref{fig:phase_eos} depicts the hadron-quark phase diagram of the employed EOS. We denote the transition density from the hadronic (mixed) to mixed (quark) phase as $\rho_{\rm tran1}$ ($\rho_{\rm tran2}$). The dependence of $\rho_{\rm tran1}$ and $\rho_{\rm tran2}$ on temperature is shown with the blue and red lines. Their dependence on $Y_e$ is relatively weak, with a difference of $\sim3\times10^{13}\,{\rm g\,cm^{-3}}$ and $\sim2\times10^{13}\,{\rm g\,cm^{-3}}$ between $Y_e=0.1$ and 0.25 for $\rho_{\rm tran1}$ and $\rho_{\rm tran2}$, respectively. This is a noteworthy difference comparing to another series of candidate hadron-quark EOSs in \cite{2011ApJS..194...39F} that have a significant dependence of the transition densities on $Y_e$ (see Fig.~5 of \cite{2011ApJS..194...39F}). For the relevant conditions at the center of PCSs, i.e. $k_{\rm B}T\sim20$\,MeV and $Y_e\sim0.25$, $\rho_{\rm tran1}$ ($\rho_{\rm tran2}$) is $\sim4.3 \,(7.7)\times10^{14}\,{\rm g\,cm^{-3}}$. 

%The original tabulated form of the employed EOS is available at CompOSE \cite{compose}. We transform it to the hdf5 format which is compatible with the EOS driver of \cite{2010CQGra..27k4103O}. The hdf5 file of the RDF 1.9 EOS is publicly available at \cite{zenodo}.

\subsection{Progenitor model}
We use the solar-metallicity 20-$M_\odot$ star of \cite{2007PhR...442..269W} (WH20) as the progenitor model for our CCSN simulations. Given that there is no accurate determination of the rotation rate of the iron core in precollapse massive stars  \cite{2005ApJ...626..350H,2021A&A...646A..19T}, we impose artificial rotation to the initial conditions with the frequently used profile \cite{1985A&A...147..161E,1997A&A...320..209Z}
\begin{equation} \label{eq:rot}
    \Omega(r) = \Omega_0 \Big[1+\Big(\frac{r}{A}\Big)^2\Big]^{-1}.
\end{equation}
Here $\Omega_0$ is the central angular speed and $r$ is the spherical radius. $A$ is a parameter characterizing the differential rotation and is chosen to be 1021\,km as \cite{2019ApJ...878...13P}. We employ five different rotation rates with $\Omega_0=0,~1,~1.5,~2~,3~,4\,$rad\,s$^{-1}$ and the models are denoted rot$i$ with $i$ indicating the magnitude of $\Omega_0$. In particular, rot0 refers to the CCSN simulation with a nonrotating progenitor. 

\subsection{CCSN simulation}
We perform CCSN simulations in 2D with the assumption of axisymmetry using the \texttt{FLASH} code \cite{2000ApJS..131..273F,2018ApJ...854...63O}. We use a cylindrical grid with the adaptive mesh refinement to cover a radius of $1.2\times10^4$\,km of the progenitor. Before the HQPT-induced collapse of the PCS takes place, the resolution of the grid is $\Delta x=$300\,m inside $\sim60$\,km and an effective angular resolution of $\Delta x/r\sim$0.6$\si{\degree}$ is employed outside $r\sim60$\,km. As observed in \cite{2022ApJ...924...38K}, we found that this leads to the decrease of central density after the bounce of the quark matter core in slowly rotating models rot0 and rot1. We adopt a higher level of refinement ($\Delta x\sim150$\,m inside $r\sim30$\,km) at a few ms before the HQPT-induced collapse to maintain the stability of the PCS. In Appendix~\ref{app:resolution} we show that GW signals of rotating models converge to the spatial resolution.
We calculate self-gravity with the multipole expansion method of \cite{2013ApJ...778..181C} with a maximum $l$ of 16 and the monopole component of the gravitational potential is modified by the Case A formalism of \cite{2006A&A...445..273M} to accommodate the general-relativistic (GR) effect.  

Neutrino transport is simulated using a two-moment scheme with the ``M1" closure and full velocity dependence for the moment equations \cite{2011PThPh.125.1255S,2013PhRvD..87j3004C,2015ApJS..219...24O,2018ApJ...865...81O}. We evolve three species of neutrinos, i.e. electron neutrino ($\nu_e$), electron antineutrino ($\bar{\nu}_e$) and heavy-flavor neutrinos ($\bar{\nu}_x$). 18 logarithmically-spaced energy bins are used to sample neutrino distribution from 0 to $\sim$ 300~MeV. Rates of neutrino-matter interactions are calculated by the \texttt{NuLib} library \cite{2015ApJS..219...24O}. Prescriptions of neutrino and hadronic matter interactions follow that of \texttt{FLASH} simulations in \cite{2018JPhG...45j4001O} except that we explicitly include inelastic neutrino-electron scattering. The calculation of rates in the mixed and quark phase follows the approximation in \cite{2009PhRvL.102h1101S} which treats quarks as nucleons and the nucleon chemical potentials are calculated as
\begin{equation} \label{eq:chem}
    \mu_{n} = \mu_u+2\mu_d; \quad   \mu_{p} = 2\mu_u+\mu_d,
\end{equation} 
where $\mu_{n(p)}$ is the chemical potential of neutrons (protons) and $\mu_{u(d)}$ is the chemical potential of up (down) quarks. The precise rate should be not important in the neutrino-trapping region where quarks exist. 

We extract the GW strain $h_+$ from our axisymmetric CCSN simulations using the standard quadrupole formula in the slow-motion and weak-field approximations \cite{1990ApJ...351..588F}
\begin{equation} \label{eq:hplus}
    h_+=\frac{3}{2}\frac{G}{Dc^4} \frac{d^2 I_{zz}}{dt^2} \sin^2 \theta_{\rm S}.
\end{equation}
Here $D$ is the distance between the source and GW detector, $I_{zz}$ is the reduced mass quadrupole moment and $\theta_{\rm S}$ is the angle between the stellar rotation axis and line of sight. Unless specified we always assume the optimal observational condition $\theta_{\rm S}=90^{\circ}$. 

\section{Results} \label{sec:result}
\subsection{CCSN dynamics} \label{ssec:dyn}
The dynamics of the stellar core in a CCSN proceeds as follows. The iron core of a massive star becomes unstable against gravity when its mass exceeds the effective Chandrasekhar limit. Dynamical collapse of the core is accelerated due to the depletion of pressure by electron captures. The collapse is halted as the core stiffens when its central density $\rho_{\rm c}$ reaches $\sim \rho_{\rm sat}$ and short-range repulsive forces between nucleons provide the pressure support. The core bounces back due to its large inertia and kinetic energy and launches a shock wave into the outer infalling matter. We denote this as the first bounce and the time $t_{1\rm b}$ is defined as the moment when the maximum entropy at the front of this shock wave reaches 3~$k_{\rm B}\,{\rm nuc}^{-1}$. The bounce shock loses its energy by disintegrating heavy nuclei to free nucleons and turns into an accretion shock in a few ms. Afterwards, the accretion shock slowly expands to $\sim$100-200 km in a few hundred ms. The stellar core, now dubbed PCS, grows in mass due to accretion. In this work, the HQPT takes place when the PCS central density $\rho_{\rm c}$ exceeds $\rho_{\rm tran1}$ and triggers the collapse of PCS shortly after $\rho_{\rm c}$ exceeds $\rho_{\rm tran2}$ (cf. Fig.~\ref{fig:phase_eos} for the transition densities). The PCS collapse, at this point is called the second collapse, is then halted by the stiffening of the quark matter EOS. The core overshoots its equilibrium state and bounces back once again. We denote this as the second bounce and define $t_{2\rm b}$ as the moment with the maximum $\rho_{\rm c}$. The second bounce shock can result in a successful CCSN explosion if it overcomes the ram pressure of the infalling stellar mantle. 

\begin{figure}[t!]
    \centering
    \includegraphics[width=0.49\textwidth]{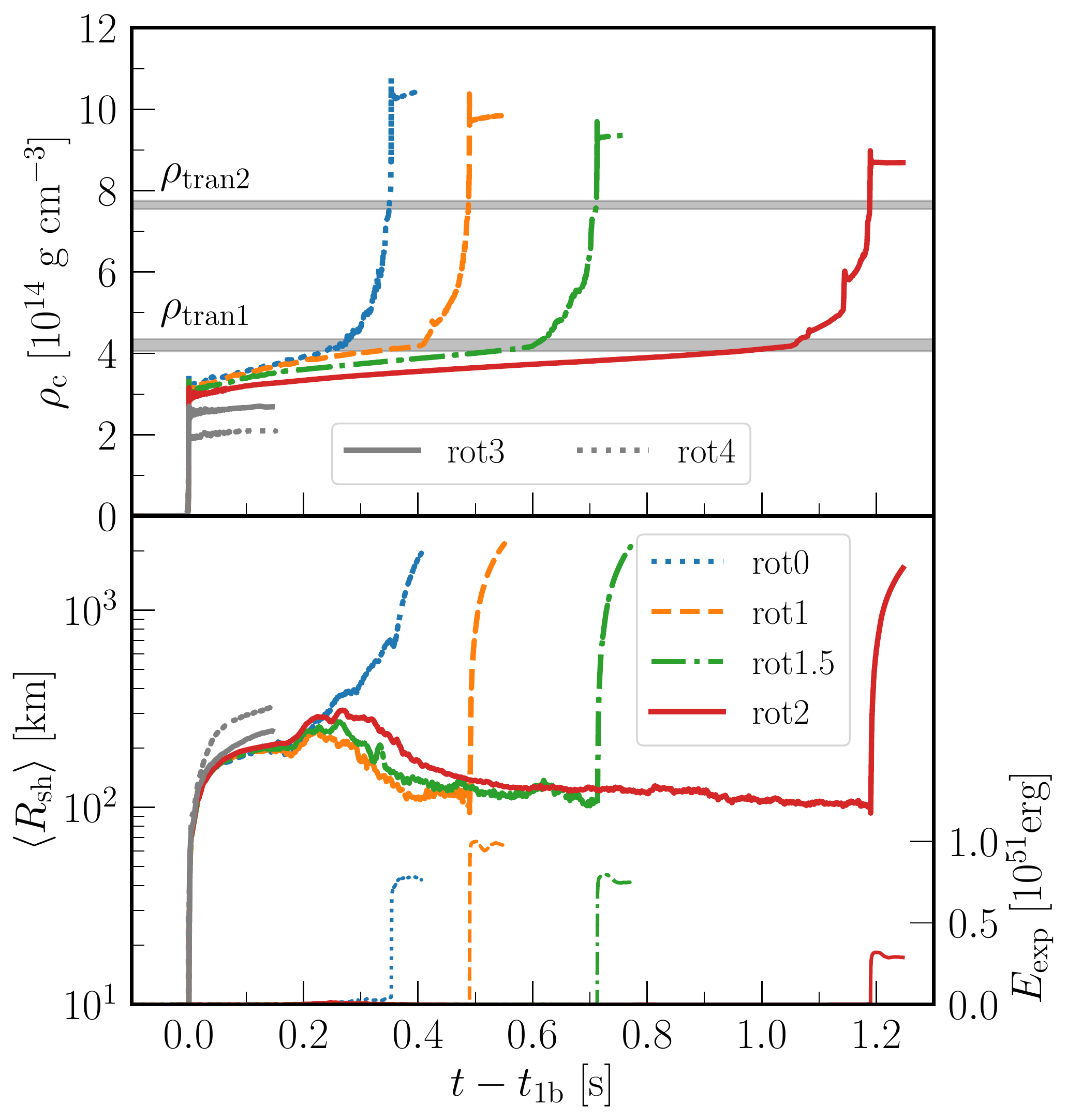}
    \caption{Time evolution of the central density (upper panel), mean shock radius (thick lines in the lower panel) and diagnostic explosion energy (thin lines in the lower panel) after the first bounce. rot$i$ refers to the model with $\Omega_0=i$\,rad\,s$^{-1}$ in Eq.~\ref{eq:rot}.  The gray shaded horizontal bands mark the transition densities from the hadron phase to the mixed phase ($\rho_{\rm tran1}$) and from the mixed phase to the quark phase ($\rho_{\rm tran2}$), taking their dependence on temperature and $Y_e$ into account.}
    \label{fig:dyn}
\end{figure}

In Fig.~\ref{fig:dyn} we plot the time evolution of the central density $\rho_{\rm c}$, mean shock radius $\langle R_{\rm sh}\rangle$, and diagnostic explosion energy $E_{\rm exp}$ after the first bounce in the models rot0...rot2. Faster rotation results in a smaller growing rate of $\rho_{\rm c}$ during the accretion phase after $t_{1\rm b}$. Also, starting from $\sim t_{\rm1b}+0.2$\,s, $\langle R_{\rm sh}\rangle$ is larger for faster rotation except for the nonrotating model rot0. The larger $\langle R_{\rm sh}\rangle$ in rot0 is likely due to the more violent neutrino-driven convection behind the shock which is stabilized by a positive angular momentum gradient in rotating models \cite{1978ApJ...220..279E,2000ApJ...541.1033F,2021ApJ...923..201E}. Unlike rotating models, $\langle R_{\rm sh}\rangle$ in rot0 does not stall at $\sim200\,$km, but expands to $\sim600$\,km before the HQPT-induced second collapse. Nevertheless, all models do not exhibit a runaway shock expansion and markedly positive $E_{\rm exp}$ before the HQPT takes place, which suggests that neutrino heating has not yet led to a successful CCSN explosion.

\begin{table*}[t!]
    \centering
    \begin{tabular}{l|cccccc}
    \hline
        Model & $t_{\rm 1b}$ & $\rho_{\rm c,max1}$ & $t_{\rm 2b}-t_{\rm 1b}$ & $M_{\rm PCS,crit}$&$\rho_{\rm c,max2}$ & $E_{\rm exp}$\\
        & [s] & [$10^{14}~{\rm g~cm^{-3}}$] & [s] & $[M_\odot]$ & [$10^{14}~{\rm g~cm^{-3}}$]& [10$^{51}$~erg]\\ \hline
        rot0   & 0.341  & 3.51 & 0.353 & 1.87 & 10.77 & 0.79 \\
        rot1   & 0.349  & 3.42 & 0.490 & 1.94 & 10.38 & 1.00\\
        rot1.5 & 0.359  & 3.32 & 0.712 & 2.02 & 9.70 & 0.80\\
        rot2   & 0.374  & 3.16 & 1.190 & 2.16 & 8.98 & 0.32\\ 
        rot3   & 0.427  & 2.70 & --- & ---& --- & ---\\
        rot4   & 0.548  & 2.08 & --- & --- & --- & ---\\ \hline
    \end{tabular}
    \caption{Major quantitative results of our hydrodynamic simulations. $t_{\rm 1(2)b}$ is the time of the first (second) bounce. $\rho_{\rm c, max1(2)}$ is the maximum central density after the first (second) core collapse. $M_{\rm PCS,crit}$ is the critical mass of the protocompact star immediately before the second collapse. Here, the protocompact star refers to the region with densities $\ge10^{11}~\rm g\,cm^{-3}$.}
    \label{tab:dyn}
\end{table*}

The collapse of PCS triggered by the HQPT is indicated by the nearly vertically increasing $\rho_{\rm c}$ after it exceeding $\rho_{\rm tran2}$ (upper panel of Fig.~\ref{fig:dyn}). Rotation significantly delays this second collapse by providing additional centrifugal support against gravity. The time interval between the first bounce and the second collapse for the model rot2 is $\sim$3.3 times longer than that for the model rot0. As a result, the central $Y_e$ prior to the second collapse is $\sim0.24$ for the model rot2, lower than $\sim0.26$ for the other models rot0...rot1.5. Note that we do not evolve the models rot3 and rot4 to the second collapse which will take much longer than 1\,s so that the assumption of axisymmetry is likely to be broken due to nonaxisymmetric instabilities. Following the second collapse and as $\rho_{\rm c}$ reaches its maximum and decreases, a second bounce shock is launched and leads to the runaway expansion of $\langle R_{\rm sh}\rangle$ (thick lines in the lower panel of Fig.~\ref{fig:dyn}) and the surge of an $\mathcal{O}(10^{51})$-erg $E_{\rm exp}$ (thin lines in the lower panel of Fig.~\ref{fig:dyn}) that signal a successful HQPT-induced explosion. The major quantitative results of the dynamics are summarized in Table~\ref{tab:dyn}. At the end of the simulation (50\,ms after $t_{\rm 2b}$), model rot1 has the largest $E_{\rm exp}$ ($\sim 10^{51}$\,erg) and while model rot2 has the smallest $E_{\rm exp}$ ($\sim 0.25\times10^{51}$\,erg). We remark that the nonmonotonic relation between $E_{\rm exp}$ and the rotation rate requires a further systematic investigation.

\begin{figure}[t!]
    \centering
    \includegraphics[width=0.49\textwidth]{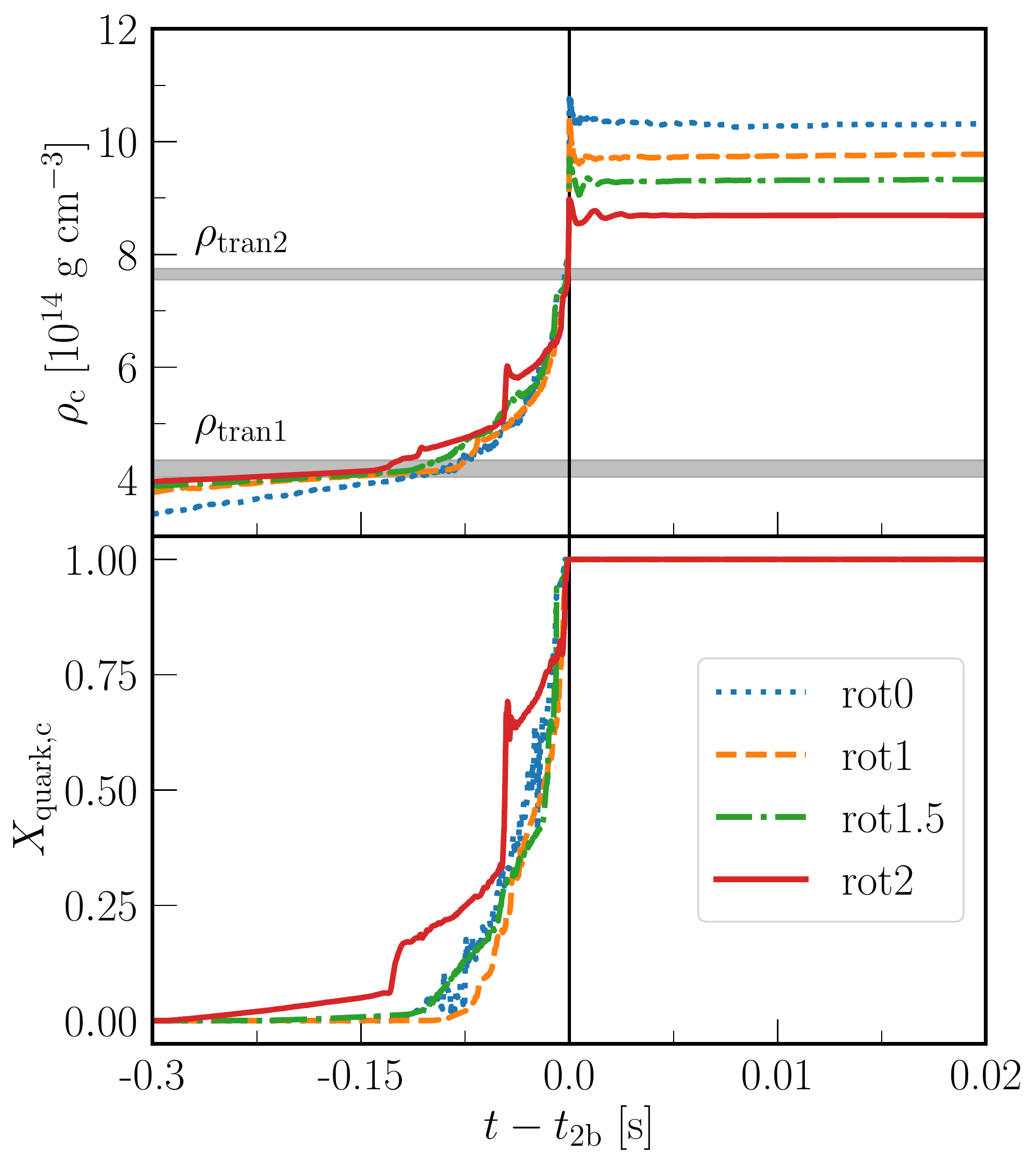}
    \caption{Time evolution of the central density (upper panel) and central quark volume fraction (lower panel) from 300\,ms before to 20\,ms after the second bounce $t_{\rm 2b}$. Note the scales are different in the epochs before and after $t_{\rm 2b}$, which is marked by the black vertical line. The gray shaded horizontal bands are the same as those in Fig.~\ref{fig:dyn}. The spurious sudden increase of $\rho_{\rm c}$ as well as $X_{\rm quark,c}$ at $t_{\rm 2b}-0.13$\,s and $t_{\rm 2b}-0.05$\,ms in the model rot2 is confined the central few km and has negligible dynamical effect on the protocompact star, cf. Fig.~\ref{fig:rm_traj}. }
    \label{fig:rhoc2b}
\end{figure}

\begin{figure*}[t!]
    \centering
    \includegraphics[width=\textwidth]{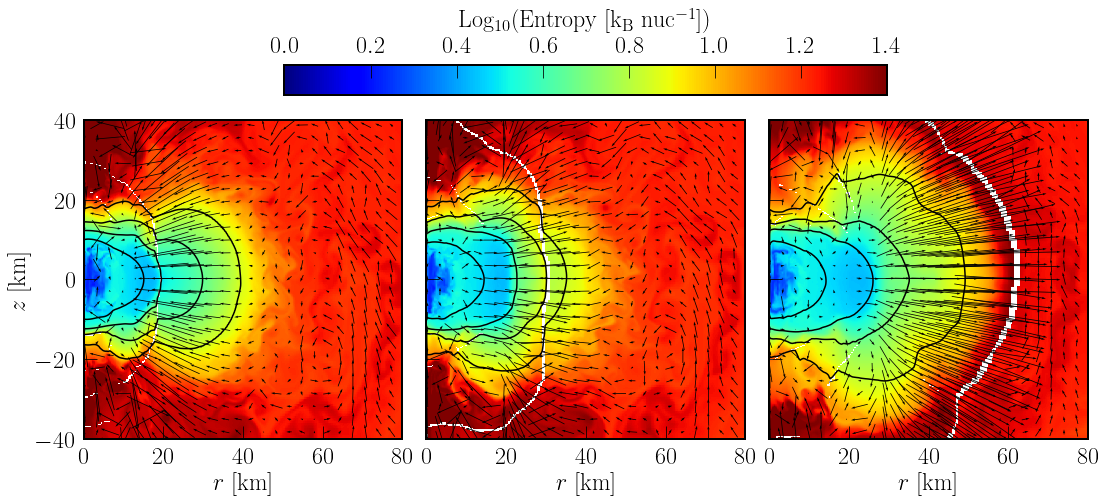}
    \caption{Snapshots illustrate the PCS dynamics and the development of the second bounce shock at $t=t_{\rm 2b}$ (left panel), $t=t_{\rm 2b}+0.3$\,ms (middle panel) and $t=t_{\rm 2b}+0.6$\,ms (right panel). $t=t_{\rm 2b}$ and $t=t_{\rm 2b}+0.6$\,ms correspond to the times of the maximum $\rho_{\rm c}$ and the following dip for model rot2 in Fig.~\ref{fig:rhoc2b}. The color maps encode the magnitude of entropy, the black curves mark the isodensity surfaces with densities of 10$^{14}$, 10$^{13}$, 10$^{12}$, 10$^{11}$\,g\,cm$^{-3}$ from the inside out, and the arrows depict velocities. The white curves mark the locus of the second shock.}
    \label{fig:rot_slice}
\end{figure*}

To have a closer look at the HQPT-induced collapse of PCS and the subsequent bounce, we plot the evolution of $\rho_{\rm c}$ and the central quark volume fraction $X_{\rm quark,c}$ from $t_{\rm 2b}-300$\,ms to $t_{\rm 2b}+20$\,ms in Fig.~\ref{fig:rhoc2b}. Note that the scales of the x-axis are different in the epochs before and after $t_{\rm 2b}$, which is marked by the black vertical line. Before $\rho_{\rm c}$ reaches $\rho_{\rm tran1}$, i.e. with purely hadronic matter at the center, $\rho_{\rm c}$ grows in an almost constant rate. As the HQPT sets in and the center enters the mixed phase, the growth rate of $\rho_{\rm c}$ increases. Eventually when $\rho_{\rm c}$ reaches $\rho_{\rm tran2}$ with purely quark matter at the center the PCS collapses dynamically and $\rho_{\rm c}$ reaches its maximum in less than 1\,ms. After $t_{\rm 2b}$, faster rotation results in a smaller $\rho_{\rm c}$ with a more prominent oscillation for a few ms, corresponding to the hydrodynamical ringdown of the strongly rotating quark matter core. Fig.~\ref{fig:rot_slice} shows the PCS dynamics and development of the second bounce shock shortly after $t_{\rm 2b}$ in the model rot2. The extreme oblateness of the PCS can be seen from the isodensity contours marked by the black lines, while the shock front has a prolate shape. These bear a close resemblance to the dynamics of the first bounce in rotating CCSNe \cite{2012PhRvD..86b4026O} .

\begin{figure}[t!]
    \centering
    \includegraphics[width=0.49\textwidth]{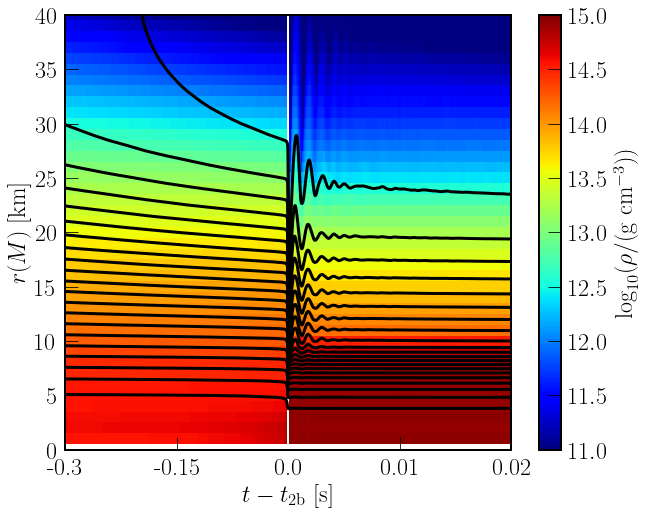}
    \caption{Black lines show the time evolution of the radius at fixed enclosed masses $r(M)$, where $M$ ranges from 0.1\,$M_\odot$ to 2.1\,$M_\odot$ with a step of 0.1\,$M_\odot$. The colormap shows the time evolution of the angular-averaged density profile. The time of the second bounce is marked by the vertical white line.}
    \label{fig:rm_traj}
\end{figure}

\begin{figure}
    \centering
    \includegraphics[width=0.49\textwidth]{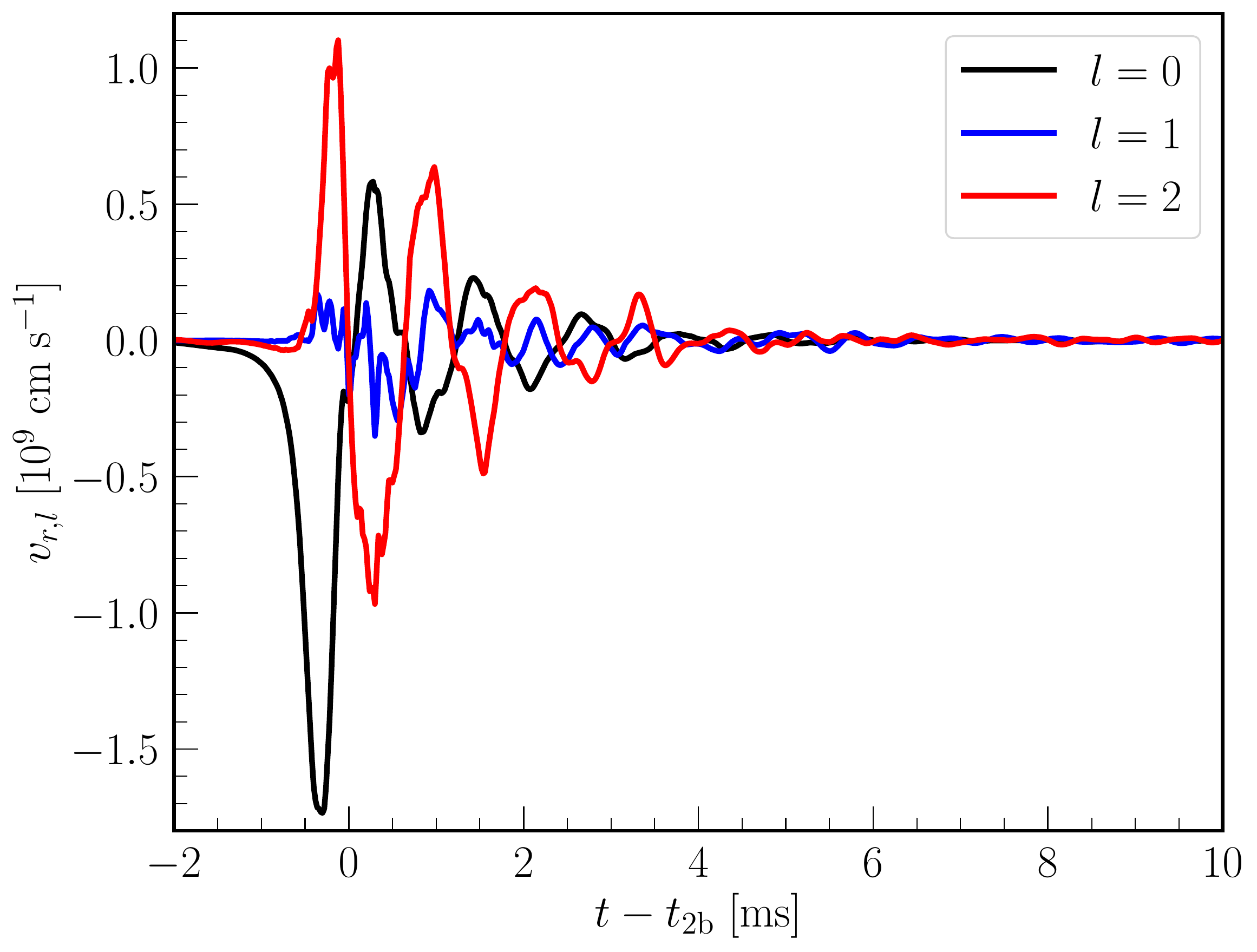}
    \caption{Coefficients with $l=0,~1,~2$ of the Legendre decomposition of the radial velocities at 10\,km. An angular resolution of $4^{\circ}$ is used for the Legendre decomposition.}
    \label{fig:vr10km}
\end{figure}

The model rot2 experiences spurious sudden increases of $\rho_{\rm c}$ and $X_{\rm quark,c}$ at 0.13\,s and 0.05\,ms before $t_{\rm 2b}$ (cf. Fig.~\ref{fig:rhoc2b}). This is likely due to the almost flat pressure-density relation in the mixed phase. However, these increases have a negligible effect on the global dynamics of the PCS core, which is depicted by Fig.~\ref{fig:rm_traj} with the time evolution of the radius at fixed enclosed masses $r(M)$, where $M$ ranges from 0.1\,$M_\odot$ to 2.1\,$M_\odot$ with a step of 0.1\,$M_\odot$. $r(M)$ shrinks slowly until the dynamical collapse signaled by the sharp decreases of $r(M)$ shortly before $t_{\rm 2b}$, which is marked by the white vertical line. After $t_{\rm 2b}$, $r(M)$ bounces up (even to a slightly larger value for $M=2.1\,M_\odot$ than that before the collapse) and then oscillates for several cycles across the following $\sim$5\,ms. The global oscillation of the PCS after $t_{\rm 2b}$ can also be seen from the time evolution of the angular-averaged density profile, which is shown with the colormap in Fig.~\ref{fig:rm_traj}. To unveil the nature of oscillations in the model rot2, we perform a Legendre decomposition of the radial velocities ($v_r$) in the PCS. Fig.~\ref{fig:vr10km} shows the lowest 3 Legendre components of $v_r$ at 10\,km from 2\,ms before to 10\,ms after $t_{\rm 2b}$. Prominent oscillation is seen for $l=0$ and 2, indicating both radial and quadrupolar nature. The $l=0$ and 2 oscillations have almost identical frequencies, similar amplitudes and similar damping timescales.

\begin{figure}
    \centering
    \includegraphics[width=0.48\textwidth]{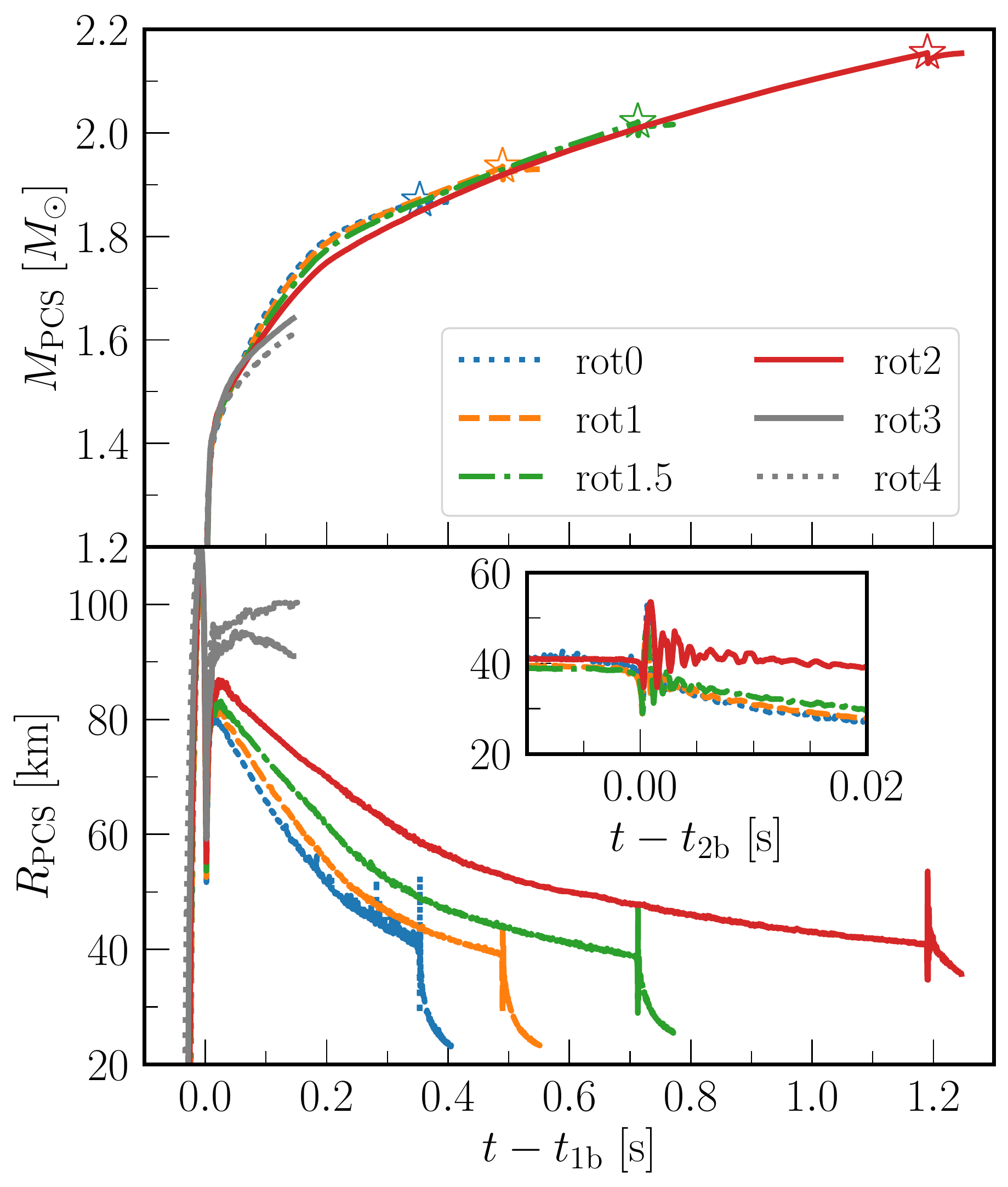}
    \caption{Time evolution of the mass (upper panel) and radius (lower panel) of the protocompact star after the first bounce. The open stars in the upper panel mark the critical $M_{\rm PCS}$ immediately before the second collapse. The inset in the lower panel exemplifies the evolution of $R_{\rm PCS}$ around the second bounce.}
    \label{fig:pcs}
\end{figure}

Rotation also impacts the time evolution of the PCS mass and radius as shown in Fig.~\ref{fig:pcs}, where the PCS surface is defined at the angular-averaged radius with a density of $10^{11}$\,g\,cm$^{-3}$. During the accretion phase after $t_{\rm 2b}$, $M_{\rm PCS}$ is slightly smaller for faster rotation which slows down the accretion and $R_{\rm PCS}$ is larger due to the centrifugal support. The second collapse is signaled by the sudden drop of $R_{\rm PCS}$, which is exemplified in the inset of the lower panel of Fig.~\ref{fig:pcs}. The critical PCS mass immediately before the second collapse ($M_{\rm PCS,crit}$, marked by the open stars in Fig.~\ref{fig:pcs} and listed in Table~\ref{tab:dyn}) is larger for faster rotation as rotational support counteracts gravity to delay the collapse. $M_{\rm PCS,crit}$ of the model rot2 is 0.29\,$M_\odot$ ($\sim$15.5\%) larger than that of the model rot0. On the other hand, $R_{\rm PCS}$ at the second collapse is almost independent of the rotation rate. The inset of the lower panel of Fig.~\ref{fig:pcs} also shows the oscillation of $R_{\rm PCS}$ after the second bounce in the rotating models.

\begin{table*}[t!]
    \centering
    \begin{tabular}{l|cccccccc}
    \hline
        Model &  $\beta_{\rm ini}$ & $\beta_{\rm PCS,1b}$ & $\beta_{\rm PCS,2c}$ & $\beta_{\rm PCS,2b}$ & $J_{\rm ini}$ & $J_{\rm PCS,1b}$ & $J_{\rm PCS,2c}$ \\
         &  [\%] & [\%] & [\%] & [\%] & [10$^{48}$erg$\cdot$s] & [10$^{48}$erg$\cdot$s] & [10$^{48}$erg$\cdot$s] \\ \hline
        rot1    & 0.04 & 1.4  & 3.0  & 4.3  &  3.30 & 0.89 & 1.84\\ 
        rot1.5  & 0.10 & 3.0  & 6.7  & 9.4  &  4.96 & 1.34 & 3.03\\ 
        rot2    & 0.18 & 5.2  & 12.2 & 15.2 &  6.61 & 1.78 & 4.62\\ 
        rot3    & 0.40 & 10.1 & ---  & ---  &  9.91 & 2.65 &--- \\ 
        rot4    & 0.71 & 14.4 & ---  & ---  & 13.20 & 3.36 &--- \\ 
        \hline
    \end{tabular}
    \caption{Quantities that measure the rotation rates of the progenitor and protocompact star. $\beta_{\rm ini(PCS)}$ is the ratio of the rotational kinetic energy to gravitational potential energy of the progenitor (protocompact star). $J_{\rm ini(PCS)}$ is the total angular momentum of the progenitor (protocompact star). The subscripts, `1b', `2c' and `2b', refer to the moments at $t_{\rm 1b}+10$\,ms, $t_{\rm 2b}-5$\,ms and $t_{\rm 2b}+10$\,ms. See Fig.~\ref{fig:beta} for the detailed time evolution of $\beta_{\rm PCS}$ and $J_{\rm PCS}$. Here, the progenitor refers to the region that has been mapped to the computational grid and the protocompact star refers to the region with densities $\ge10^{11}~\rm g\,cm^{-3}$. }
    \label{tab:rot}
\end{table*}

\begin{figure}[t!]
    \centering
    \includegraphics[width=0.49\textwidth]{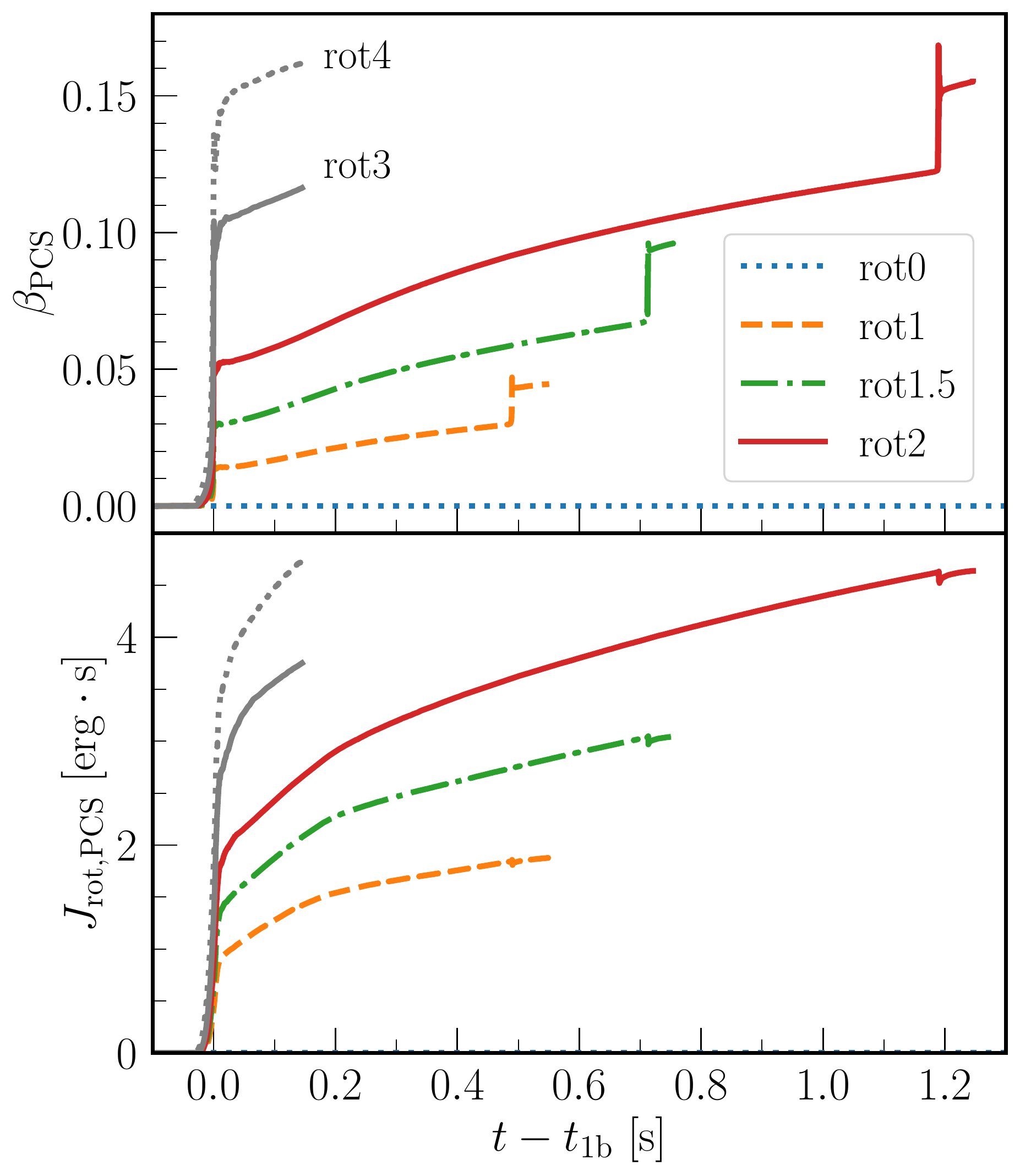}
    \caption{Time evolution of the parameter $\beta$ (upper panel) and the total angular momentum (lower panel) of the protocompact star after the first bounce.}
    \label{fig:beta}
\end{figure}
    
The PCS rotation is greatly accelerated during the dynamical collapse and postbounce accretion phase because more angular momentum gets concentrated in it. Fig.~\ref{fig:beta} depicts this spin-up process with the time evolution of $\beta_{\rm PCS}$ (upper panel) and $J_{\rm PCS}$ (the PCS total angular momentum, lower panel) after the first bounce. Here $\beta$ is the ratio of the rotational kinetic energy $T$ to gravitational potential energy $|W|$, and $|W|$ is calculated with the approximate GR gravitational potential with the Case A formula \cite{2006A&A...445..273M}. A sudden increase of $\beta_{\rm PCS}$ is seen at both the first and second collapse as $R_{\rm PCS}$ decreases, while $\beta_{\rm PCS}$ grows with a nearly constant rate in the postbounce accretion phases. Values of $\beta_{\rm PCS}$ and $J_{\rm PCS}$ at some critical time points are listed in Table~\ref{tab:rot}. It is interesting to note that after the second bounce, the model rot2 (rot1.5) reaches similar $\beta_{\rm PCS}$ and $J_{\rm PCS}$ than those of the model rot4 (rot3) after the first bounce, while the initial central angular velocity $\Omega_{\rm0}$ is smaller by a factor of 2. This demonstrates that the HQPT-induced collapse can yield a significant PCS rotation rate that is unlikely in a single collapse scenario of the canonical CCSNe with a normal progenitor rotation rate \cite{2005ApJ...626..350H}. The $\beta_{\rm PCS}$ after the second bounce in the model rot2 ($\sim15\%$) is greater than the threshold value for the secular bar-mode instability (in Newtonian theory $\beta_{\rm sec}=14\%$  \cite{2001AIPC..575..246L}, while GR can lead to a smaller $\beta_{\rm sec}$ \cite{2003LRR.....6....3S}). Persistent GW emission can arise from nonaxisymmetric instabilities and warrants further investigation.

\subsection{Gravitational waves} \label{ssec:gw}

\begin{table*}[t!]
    \centering
    \begin{tabular}{l|cccc|cc}
    \hline
        Model & $h_{\rm pos1}D$ & $h_{\rm neg1}D$ & $h_{\rm pos2}D$ & $h_{\rm neg2}D$ & $L_{\bar{\nu}_e,\rm max}$ & $\langle E_{\bar{\nu}_e}\rangle_{\rm max}$\\ 
          & $[{\rm cm}]$ & $[{\rm cm}]$ & $[{\rm cm}]$ & $[{\rm cm}]$ & $[10^{51}~\rm erg~s^{-1}]$ & $\rm[MeV]$\\ \hline
        rot0 & 5 & -6 & 515 & 548 & 556 &  28.5\\
        rot1 & 30 & -74 & 340 & -1068 & 695 & 34.5\\
        rot1.5 &58 & -145 & 948 & -3190 & 564 & 31.6\\
        rot2 & 84 & -219 & 1074 & -1836 & 267 & 25.2\\
        rot3 & 96 & -215 &  --- & --- & ---& --- \\
        rot4 & 61 & -146 &  --- & --- & ---& --- \\ \hline
    \end{tabular}
    \caption{Quantities about the main features of the multimessenger signals. $h_{\rm pos1(2)}$ and $h_{\rm neg1(2)}$ are the GW strain of the positive and negative peaks around $t_{\rm 1(2)b}$. $D$ is the distance. Note that for the model rot0, we take the extreme amplitudes of $h_+$ around $t_{\rm 1(2)b}$ as $h_{\rm pos1(2)}$ and $h_{\rm neg1(2)}$. $L_{\bar{\nu}_e,\rm max}$ and $\langle E_{\bar{\nu}_e}\rangle_{\rm max}$ are the maximum angular-averaged luminosity and mean energy of electron antineutrino ($\bar{\nu}_e$) of the neutrino burst associated with the second bounce shock.}
    \label{tab:messenger}
\end{table*}

\begin{figure*}[t!]
    \centering
    \includegraphics[width=0.95\textwidth]{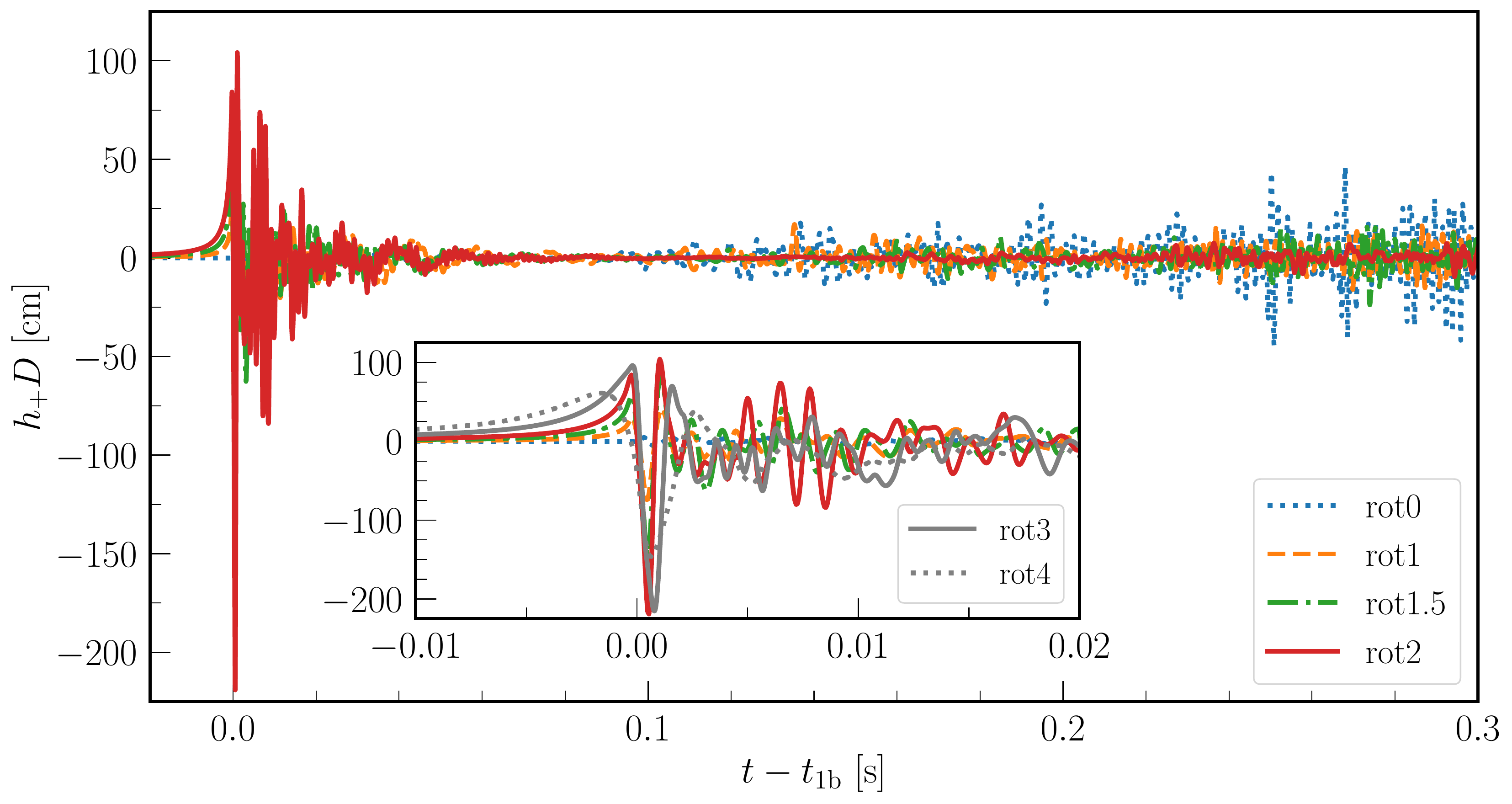}
    \caption{Time evolution of the GW strain from 0.02\,s before to 0.3\,s after the first bounce in the models rot0$\ldots$rot2. The inset exemplifies the episode between $t_{\rm1b}-0.01$\,s and $t_{\rm1b}+0.02$\,s and includes the models rot3 and rot4.}
    \label{fig:gw1}
\end{figure*}

In Fig.~\ref{fig:gw1}, we plot the GW strain ($h_{+}D$) as a function of time from 0.02\,s before to 0.3\,s after the first bounce in the models rot0$\ldots$rot2. The inset in Fig.~\ref{fig:gw1} more clearly illustrates the waveforms around $t_{\rm 1b}$, including also those of the models rot3 and rot4. The bounce signals exhibit the generic features routinely found for rotating CCSN models, cf. \cite{2007PhRvL..98y1101D}. A prominent positive (negative) peak is found before (after) $t_{\rm 1b}$ with a magnitude of $h_+D\sim\mathcal{O}(100)$~cm, which is designated as $h_{\rm pos1}$ ($h_{\rm neg1}$) with its magnitude listed in Table~\ref{tab:messenger}. These reflect the markedly quadrupolar deformation of the stellar core due to rotation, while the magnitude of its quadrupole moment rapidly changes during the dynamical collapse and bounce. An important fact is that the magnitudes of $h_{\rm pos1}$ and $h_{\rm neg1}$ increase for larger rotation rates from rot1 to rot2, but saturate and even decrease as the rotation rate goes beyond rot2. An episode of GW emission follows $h_{\rm neg1}$ for about 50\,ms with a frequency of several hundred Hz, corresponding to the hydrodynamical ringdown of the fast rotating PCS. Afterwards, GW emission subsides for about 50\,ms until its resurgence at $\sim t_{\rm 1b}+100$\,ms due to PCS oscillations during the accretion phase. $h_+$ is smaller for a larger rotation rate, likely due to the stabilization of convection by a positive angular momentum gradient according to the Solberg-H{\o}iland stability criterion \cite{1978ApJ...220..279E,2000ApJ...541.1033F}. The GW signals around the bounce and during the accretion phase in rotating CCSNe have been studied in great detail, and we refer the interested readers to \cite{2014PhRvD..90d4001A,2017PhRvD..95f3019R,2021ApJ...914...80P} for the utilization of the signals to measure the rotation rate and progenitor compactness, and to constrain the nuclear matter EOS.

\begin{figure*}[t!]
    \centering
    \includegraphics[width=0.95\textwidth]{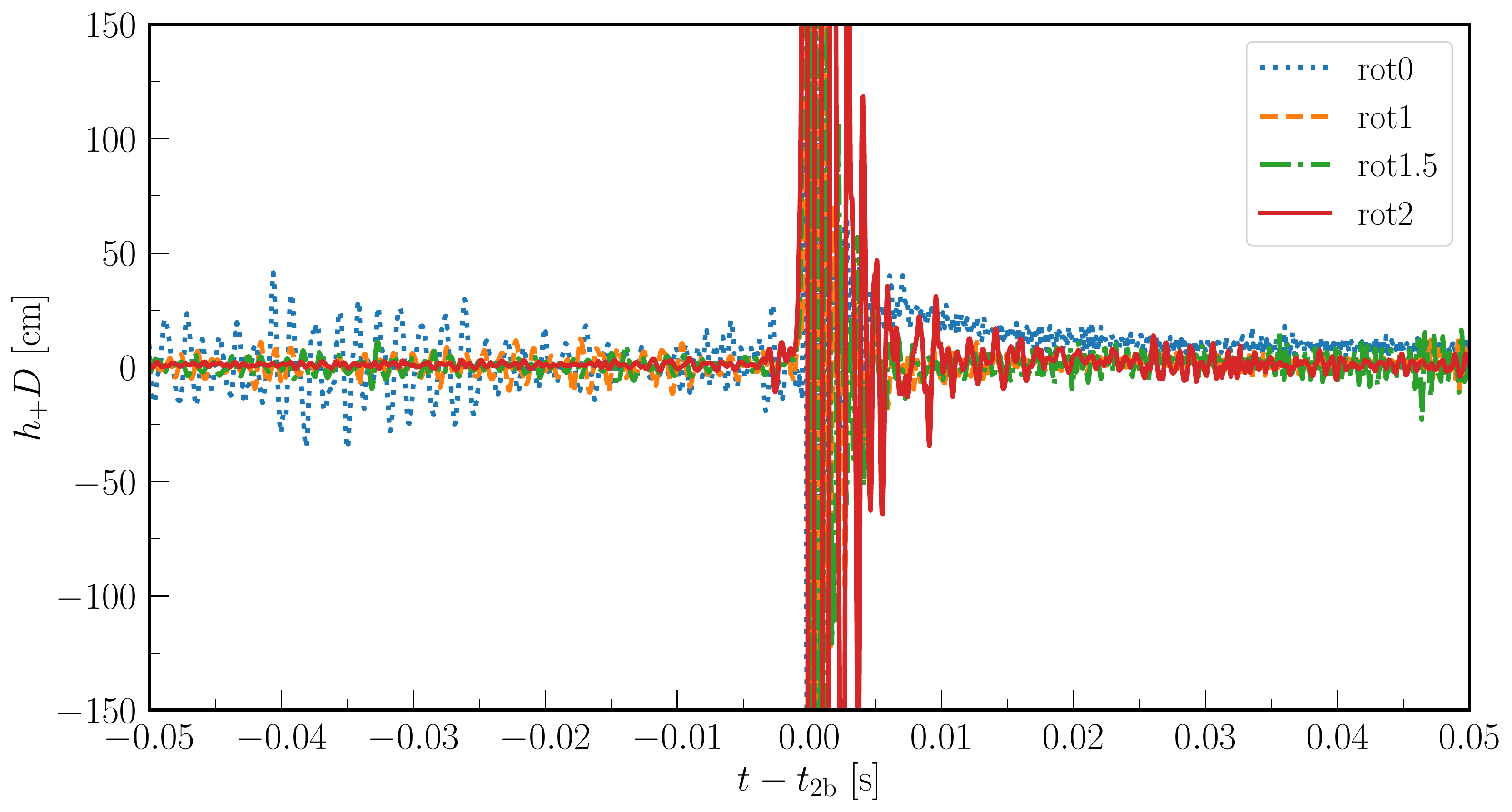}
    \caption{Same as Fig.~\ref{fig:gw1}, but for the time evolution of the GW from 0.05\,s before to 0.05\,s after the second bounce in the models rot0$\ldots$rot2.}
    \label{fig:gw2}
\end{figure*}

\begin{figure}[t!]
    \centering
    \includegraphics[width=0.49\textwidth]{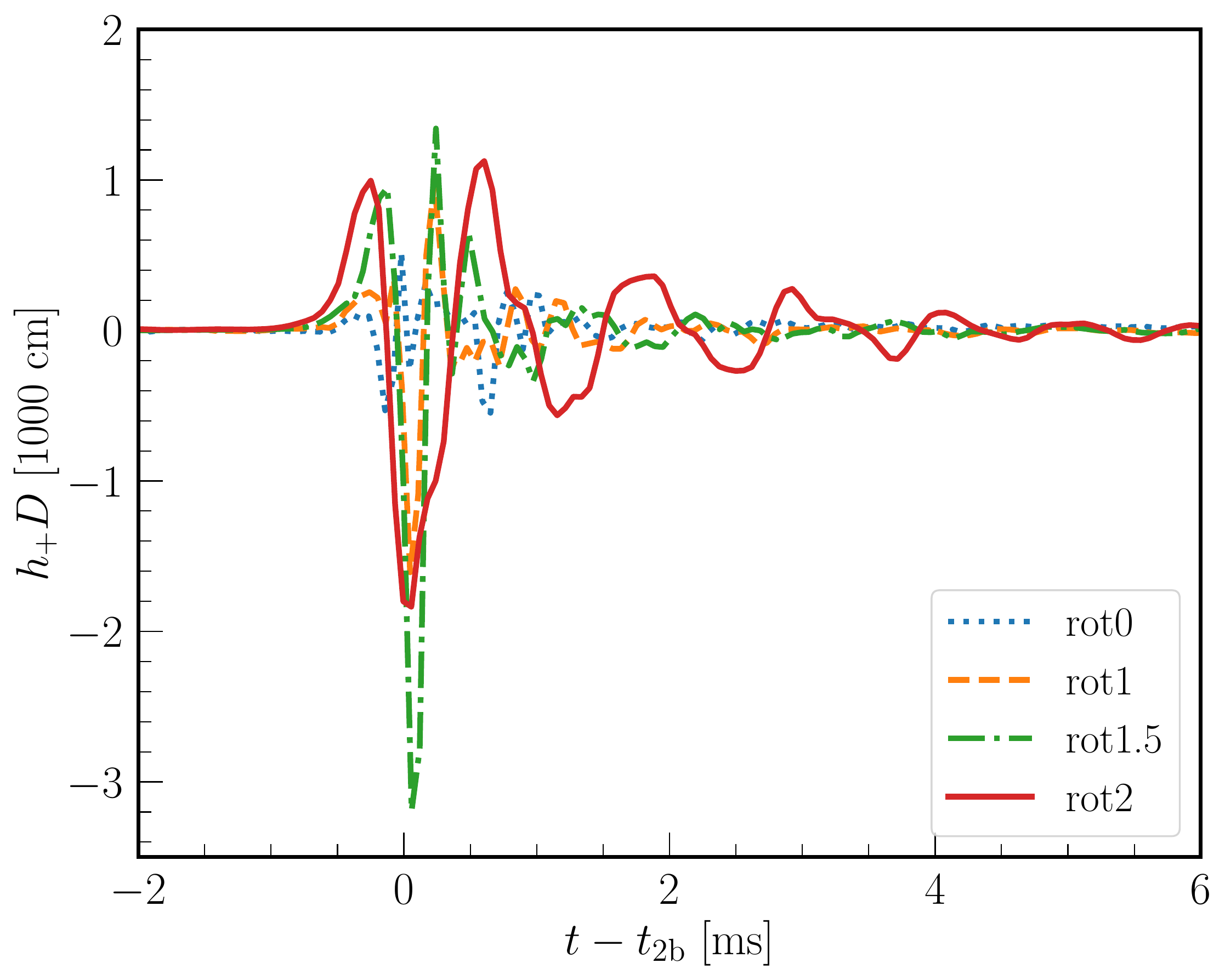}
    \caption{Time evolution of the GW strain from 2\,ms before to 10\,ms after the second bounce in the models rot0$\ldots$rot2. This figure focuses on the episode in Fig.~\ref{fig:gw2} with saturated magnitudes. Note the different $y$-axis scale compared to that in Fig.~\ref{fig:gw2}. }
    \label{fig:gw2b}
\end{figure}

Fig.~\ref{fig:gw2} depicts the GW waveforms of the models rot0$\ldots$rot2 in a time window of 0.1\,s centered at $t_{\rm 2b}$. Before the second collapse, the magnitude of $h_+D$ is $\mathcal{O}$10\,cm and smaller for a larger rotation rate. The second dynamical collapse and bounce result in a burst of GW emission (illustrated in Fig.~\ref{fig:gw2b}) that resembles that during the first collapse and bounce but with a much larger $h_+D$. The prominent positive (negative) peak before (after) $t_{\rm 2b}$ is dubbed $h_{\rm pos2}$ ($h_{\rm neg2}$) with the magnitudes listed in Table~\ref{tab:messenger}. The magnitude of $h_+D$ at these peaks is $\mathcal{O}(1000)$\,cm for the rotating models, which is an order of magnitude larger than that of $h_{\rm pos1}$ and $h_{\rm neg1}$. It can be understood by the approximate relation between the GW amplitude $\Delta h=h_{\rm pos}-h_{\rm neg}$ and the properties of the rotating core derived in \cite{2017PhRvD..95f3019R,2021ApJ...914...80P}:
\begin{equation} \label{eq:dh}
    \Delta h \sim \frac{G}{c^4DMR^2} J^2. 
\end{equation}
Here, $M$, $R$, and $J$ are the mass, radius, and total angular momentum of the rotating PCS core. $M$ and $J$ are about a factor of 2 larger at the second bounce than that at the first bounce, while $R$ is about half. By Eq.~\ref{eq:dh} a factor of $\sim$8 is expected for the enlargement of $\Delta h$ at the second bounce, roughly in agreement with the result of simulations. In Appendix~\ref{app:resolution} we show the convergence of the GW signal at the second bounce for the model rot2 with different spatial resolutions.

The nonrotating model also generates a strong GW pulse during this episode with a maximum $h_+D$ of $\sim700$\,cm, which is a factor of 2-4 larger than that in \cite{2020PhRvL.125e1102Z,2022ApJ...924...38K}. However, the ratio of this peak magnitude to the mean GW amplitude before the second collapse ($\sim30$\,cm) is about $25$, which is compatible with the published results. Yet this relation for the GW amplitudes in nonrotating models needs to be checked in a systematic way for different progenitor models and EOSs.

After the second bounce, the GW amplitude drops to $h_+D\sim\mathcal{O}(10)$\,cm after about 5\,ms. The slowly varying deviation of $h_+$ from 0 for the model rot0 corresponds to the ejection of neutrino-heated materials behind the expanding accretion shock (cf. Fig.~\ref{fig:dyn}). We also note a dramatic increase of the peak GW frequency after the second bounce, similar to the results in \cite{2020PhRvL.125e1102Z,2022ApJ...924...38K}.

\begin{figure}
    \centering
    \includegraphics[width=0.49\textwidth]{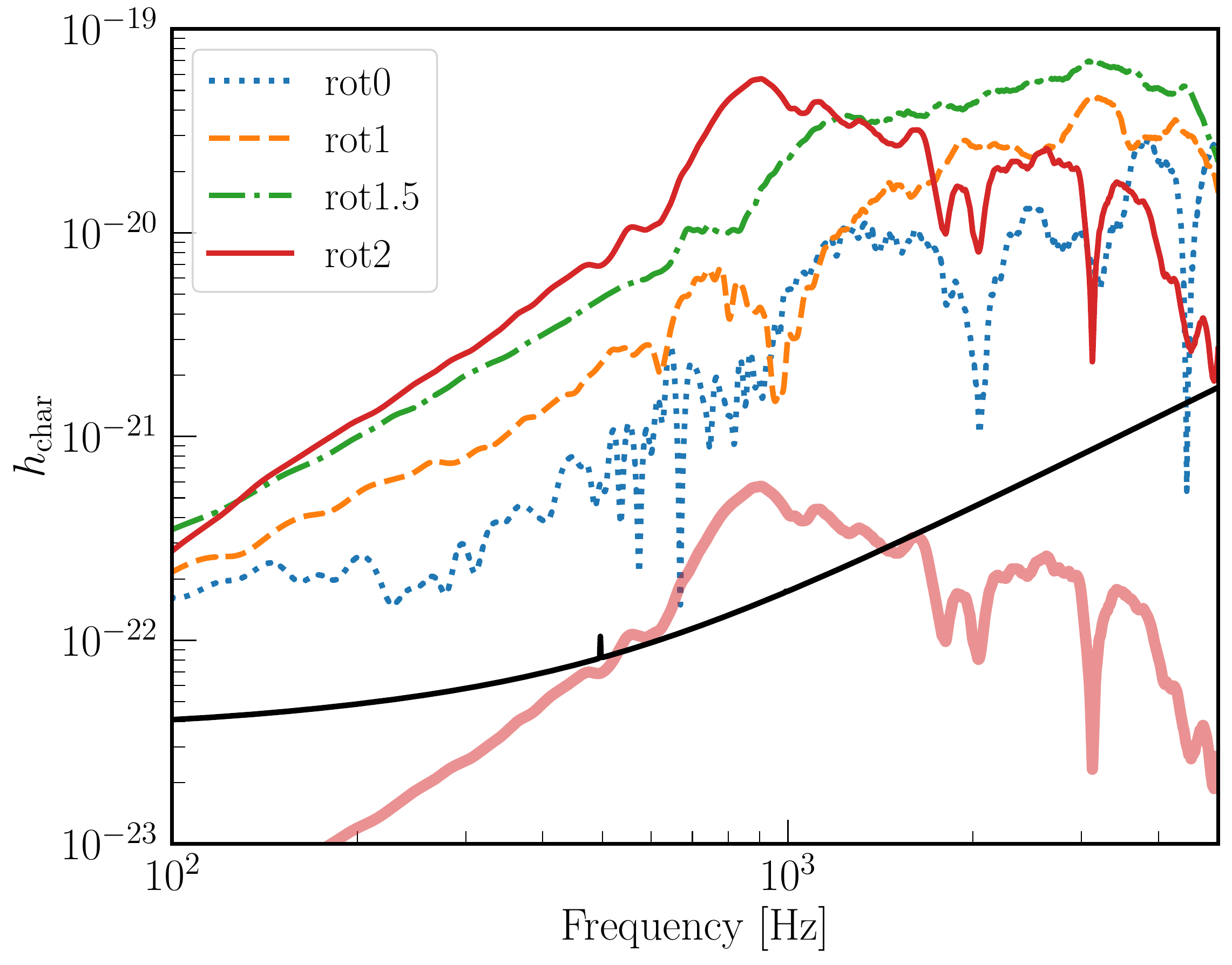}
    \caption{Characteristic GW strain $h_{\rm char}$ for the models rot0$\ldots$rot2 in the 100-ms Hann window centered at $t_{\rm 2b}$ (cf. Fig.~\ref{fig:gw2}). The black solid line is the amplitude spectral density of the detector noise multiplied by the square root of frequency for Advanced LIGO at its design sensitivity. The $h_{\rm char}$ spectra assume a source distance of 10\,kpc except the thick red line whose distance is 1\,Mpc.}
    \label{fig:hchar}
\end{figure}
To assess the GW detectability, we calculate the characteristic GW strain $h_{\rm char}$ from its spectral energy density following Ref.~\cite{1998PhRvD..57.4535F}. After some simple math, the formula reads
\begin{equation} \label{eq:hchar}
    h_{\rm char}(f)=\sqrt{\frac{32}{15}} f|\tilde{h}_+|,
\end{equation}
where $\tilde{h}_+$ is the Fourier transform of $h_+$. Fig.~\ref{fig:hchar} shows the spectra of $h_+$ for the models rot0$\ldots$rot2 in the 100-ms Hann window centered at $t_{\rm 2b}$ compared to the sensitivity spectrum of Advanced LIGO (solid black curve, \cite{LIGO_v5}). The distance is assumed to be 10\,kpc except the thick red curve, which assumes a distance of 1\,Mpc for the model rot2. The models rot0, rot1.0, and rot1.5 exhibit a broad peak at several kHz, similar to that found in nonrotating CCSN models \cite{2020PhRvL.125e1102Z,2022ApJ...924...38K}. The model rot2 has a narrower peak centered at $\sim800$\,Hz, which corresponds to the prominent oscillations following $h_{\rm neg2}$. Note that the secondary peaks at $\sim$1000\,Hz for the models rot0, rot1.0, and rot1.5 originate from GW emission during the 50-ms episode before $t_{\rm 2b}$. The GW signals are likely to be detectable for all models in the Milky Way Galaxy by Advanced LIGO, while the model rot2 may allow detection for an event 1-Mpc away.

\subsection{Neutrino signals} \label{ssec:neu}

\begin{figure}
    \centering
    \includegraphics[width=0.49\textwidth]{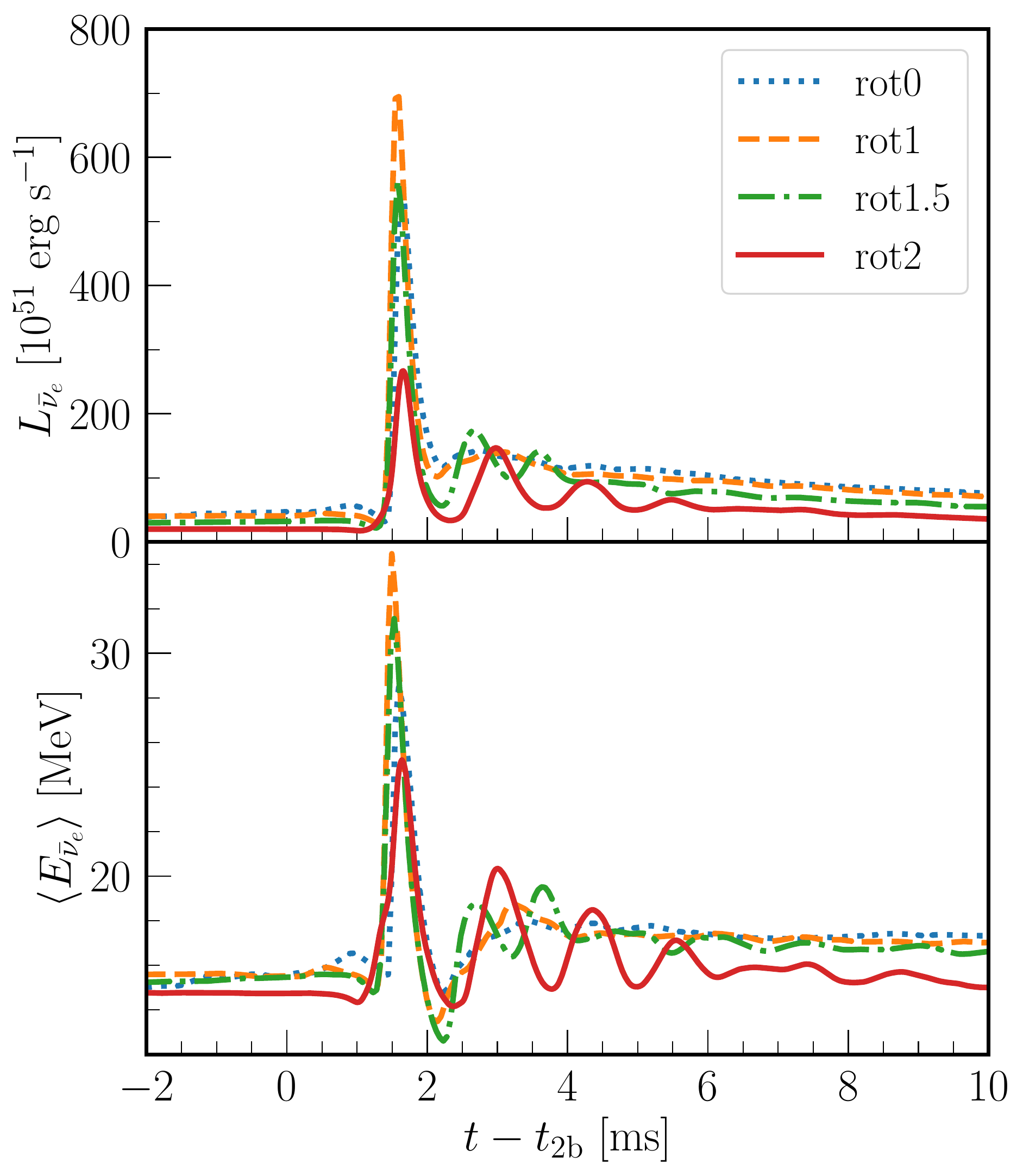}
    \caption{Time evolution of the angular-averaged luminosity (upper panel) and mean energy (lower panel) of electron antineutrinos around the second collapse and bounce.}
    \label{fig:lnua}
\end{figure}

It has been discovered by 1D \cite{2009PhRvL.102h1101S,2018NatAs...2..980F} and 2D \cite{2020PhRvL.125e1102Z,2022ApJ...924...38K} nonrotating CCSN simulations that the breakout of the second bounce shock at the neutrinosphere results in a ms $\bar{\nu}_e$-rich neutrino burst. Here we focus on the impact of rotation on this neutrino signature of the HQPT. In Fig.~\ref{fig:lnua}, we plot the angular-averaged luminosity curve $L$ and mean energy $\langle E\rangle$ of $\bar{\nu}_e$ from 2\,ms before to 10\,ms after the second bounce for the models rot0$\ldots$rot2. The quantities are extracted at a radius of 400\,km so that the neutrino burst associated with the second bounce peaks at $\sim t_{\rm 2b}+1.5$\,ms. Interestingly for the rotating models rot1.5 and rot2, oscillations in both $L_{\bar{\nu}_e}$ and $\langle E_{\bar{\nu}_e}\rangle$ with a period of $\sim$1.2-1.3\,ms are clearly seen following the first peak for $\sim5$\,ms. In Appendix~\ref{app:resolution}, we show the convergence of the neutrino signal with different spatial resolutions and the visual correlation of the oscillations in the multimessenger signals and PCS modulo the time shift. Another important fact is that the order of the maximum $L_{\bar{\nu}_e}$ and $\langle E_{\bar{\nu}_e}\rangle$ (cf. Table~\ref{tab:messenger}) is the same as that of $E_{\rm exp}$, with rot1\,$>$\,rot1.5\,$\gtrsim$\,rot0\,$>$\,rot2. The potential relation of $L_{\bar{\nu}_e,\rm max}$ and $\langle E_{\bar{\nu}_e}\rangle_{\rm max}$ with $E_{\rm exp}$ is another interesting extension of our work to be explored with a systematic set of simulations.

\begin{figure}
    \centering
    \includegraphics[width=0.49\textwidth]{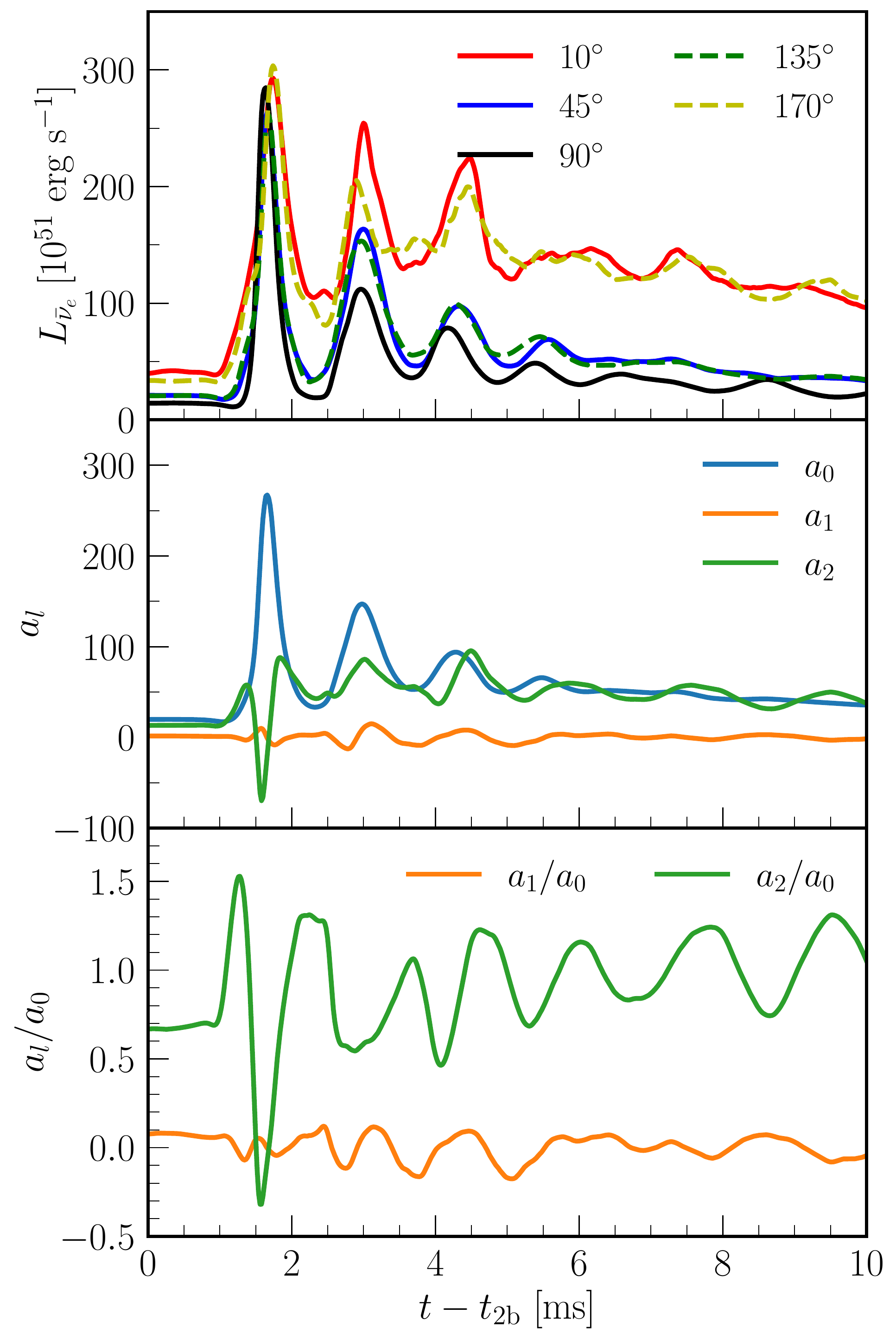}
    \caption{Time evolution of the angular dependent luminosity of electron antineutrinos (top panel), the coefficients $a_l$ of its Legendre decomposition (middle panel) and $a_l/a_0$ (bottom panel). An angular bin of $20^{\circ}$ is used in the upper panel to reduce numerical noises. An angular resolution of $4^{\circ}$ is used for the Legendre decomposition.}
    \label{fig:lnua_angular}
\end{figure}

Rotation also leads to directional variation of the neutrino signals. A smaller neutrino luminosity is expected at the equator than that at the poles because the more extended density profile due to rotational support delays and suppresses the leakage of neutrinos \cite{2006ApJ...644.1063D}. The upper panel of Fig.~\ref{fig:lnua_angular} shows the angular dependent luminosity of $\bar{\nu}_e$ at five $20^{\circ}$ angular bins. A larger $L_{\bar{\nu}_e}$ is observed for a smaller angle with respect to the poles.  Oscillations in $L_{\bar{\nu}_e}$ are seen for all directions. We plot the coefficients $a_l$ of a Legendre decomposition and $a_l$ divided by the monopole luminosity $a_0$ in the middle and lower panels of Fig.~\ref{fig:lnua_angular}, respectively. In addition to the quadrupolar signal already present before the second collapse (due to the rotation), we see an oscillatory $a_2/a_0$, indicating the quadrupolar oscillation of the PCS. Following the second collapse the baseline $a_2/a_0$ seems to also be increasing, owing to the stronger deformation of the rapidly rotating quark core.

Lastly, we remark on the difference of the $L_{\bar{\nu}_e}$ oscillation found in this work with that of \cite{2021ApJ...911...74Z}. Reference~\cite{2021ApJ...911...74Z} simulated a set of failed CCSN models in which the second bounce shock is unable to overcome the ram pressure of the envelope. After the shocked materials fall back, the PCS then oscillates for tens of ms due to its excess kinetic energy. The PCS oscillations there are purely radial. In this work, the PCS oscillations stem from the hydrodynamical ringdown of the rapid rotating core and have both radial and quadrupolar nature. In this work, since the HQPT leads to successful explosions, there is no excess kinetic energy in the PCS to give such persistent radial oscillations as \cite{2021ApJ...911...74Z}.

\section{Conclusions} \label{sec:conclu}
We have explored the impact of rotation on the multimessenger signatures of a HQPT in CCSNe. We performed a suite of CCSN simulations using \texttt{FLASH}, starting from the progenitor model WH20 with several different initial rotation rates. The HQPT is prescribed as first-order in the RDF1.9 EOS \cite{2021PhRvD.103b3001B}. The PCS born in CCSNe collapses shortly after its center transits to purely quark matter and rotation delays this second collapse significantly due to the centrifugal effect. The second bounce shock results in successful explosions in all models with the HQPT, while the diagnostic explosion energy $E_{\rm exp}$ has a nonmonotonic trend with the increasing rotation rate (cf.~Table~\ref{tab:dyn}). 

With rotation, the second collapse and bounce give rise to strong GW emission with an amplitude 10 times larger than that generated by the first collapse and bounce. GW frequency during this phase is at several kHz except for the fastest model with the HQPT (rot2). The PCS pulsation during the hydrodynamical ringdown phase results in a peak GW frequency of $\sim800$\,Hz. The GW signal of the model rot2 may allow a detection at a distance of 1-Mpc by Advanced LIGO. Meanwhile, the second collapse leads to a significant acceleration of the PCS rotation, with $\beta_{\rm PCS}>\beta_{\rm sec}=0.14$ in the model rot2. Persistent GW emission may arise from secular nonaxisymmetric instabilities due to the large $\beta_{\rm PCS}$. However, we remark that here $\beta_{\rm PCS}$ is calculated with an approximate GR gravitational potential \cite{2006A&A...445..273M} and its reliability needs to be verified in real GR hydrodynamic calculations.

The breakout of the second bounce shock at the neutrinosphere produces a prominent $\bar{\nu}_e$-rich neutrino burst. The peak luminosity of $\bar{\nu}_e$ in this burst has the same hierarchy as $E_{\rm exp}$ with respect to the rotation rate. The postbounce pulsation in the rapidly rotating models (rot1 and rot2) leads to oscillations of the neutrino signal after the breakout neutrino burst. The frequency of this oscillation is locked with the GW emission at the same time. The directional decomposition of the neutrino luminosity shows the quadrupolar nature of the oscillations. A joint detection of the GW and neutrino signals may help us to infer the occurrence of HQPT in PCS and to measure the PCS rotation rate.

The main caveat for this work is the assumption of axisymmetry that completely ignores the possibility of nonaxisymmetric hydrodynamic instabilities developing due to the fast rotation. Early studies on rapidly rotating relativistic stars suggested that the dynamical $m=1$ instability occurs for a large $\beta$ of $\sim$0.25 \cite[see e.g. ][]{2007CQGra..24S.171M}, which is not reached by our fastest rotating model. Recent CCSN simulations have found that nonaxisymmetric instabilities may set in for a much lower value of $\beta$, though subject to the methods and resolution employed by simulations. \cite{2021MNRAS.502.3066S} found nonaxisymmetric instabilities for $\beta\simeq0.07$ which lead to the decrease of $\beta$ and the increase of $\rho_\mathrm{c}$. Magneto-hydrodynamics simulations in \cite{2022arXiv221005012B} saw a even lower $\beta$ of $\sim0.03$ for nonaxisymmetric instabilities. Nonetheless, if a HQPT occurs when the PCS retains a finite $\beta$, we expect similar imprints in the multimessenger signals associated with the HQPT-induced collapse of PCS as those found in this work.

Our work has demonstrated the main features of the multimessenger signals of rotating CCSNe with the occurrence of a HQPT. All the multimessenger data are publicly available at \cite{zenodo} to be used for the prospect of detection. We expect further systematic investigation to unveil the quantitative relations between the observable and simulation inputs, such as the progenitor mass and initial rotation rate, as well as the prescription of the HQPT. Moreover, the assumption of axisymmetry during the accretion phase after the first bounce should be checked with three-dimensional and perhaps fully GR simulations \cite{2021MNRAS.502.3066S}.

\begin{acknowledgments}
We thank Sean Couch for \texttt{FLASH} code development and Dong Lai for useful discussion about nonaxisymmetric instabilities and comments on the manuscript. This research made use of \texttt{FLASH} \cite{2000ApJS..131..273F}, \texttt{yt} \cite{yt}, \texttt{Numpy} \cite{numpy}, \texttt{Matplotlib} \cite{matplotlib} and \texttt{CompOSE} \cite{compose,2022arXiv220303209T}. This work is supported by the China Postdoctoral Science Foundation (2022M712082) and the Swedish Research Council (Project No. 2020-00452). The simulations were partly run on the Siyuan-1 cluster supported by the Center for High Performance Computing at Shanghai Jiao Tong University and resources provided by the Swedish National Infrastructure for Computing (SNIC) at PDC and NSC partially funded by the Swedish Research Council through grant agreement No. 2016-07213.
\end{acknowledgments}

\renewcommand\thefigure{\thesection.\arabic{figure}}   
\appendix

\section{Convergence test of spatial resolution} \label{app:resolution}
\setcounter{figure}{0}   
\begin{figure}[t!]
    \centering
    \includegraphics[width=0.49\textwidth]{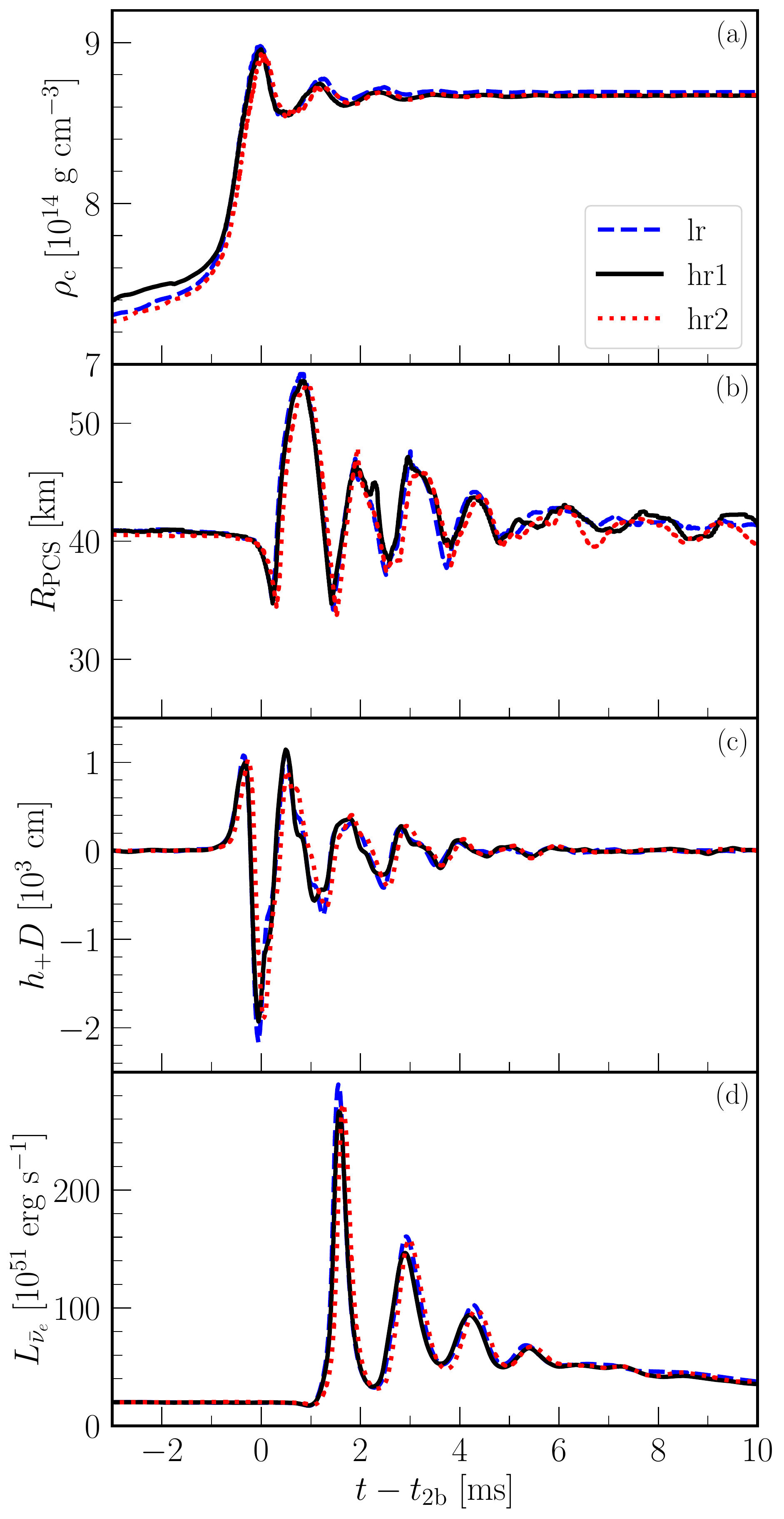}
    \caption{Time evolution of the central density (panel a), PCS radius (panel b), GW strain (panel c) and angular-averaged luminosity of electron antineutrinos (panel d) in the model rot2 with different spatial resolutions. The setup of `hr1' is used in the main text. See text for the detailed description of each simulation.}
    \label{fig:res}
\end{figure}
Here, we present a convergence test of the spatial resolution for the model rot2 (Fig.~\ref{fig:res}). `lr' refers to the simulation with the same highest level of refinement throughout the run, i.e. $\Delta x=300$\,m inside $\sim60$\,km. The `hr1' and `hr2' simulations add a higher level of refinement ($\Delta x=150$\,m) for the PCS inside $\sim30$\,km starting from $\sim4$\,ms and $\sim$140\,ms before $t_{\rm 2b}$. In Fig.~\ref{fig:res} we compare the time evolution of the central density (panel a), mean PCS radius (panel b), GW strain (panel c), and angular-averaged luminosity of $\bar{\nu}_e$ in these 3 simulations from $t_{\rm 2b}-$3\,ms to $t_{\rm 2b}+10$\,ms. It shows an excellent convergence of the multimessenger signals of the HQPT in a rapid rotating CCSN model.

\newpage

\bibliography{the_bib}% Produces the bibliography via BibTeX.

\end{document}